\documentclass[12pt,a4paper]{article}

\usepackage[bulletsep]{collref}
\usepackage{amsmath,amssymb,graphicx}
\usepackage{pst-all}
\usepackage{bbm}
\usepackage{epstopdf}

\makeatletter \@addtoreset{equation}{section} \makeatother

\makeatletter
\let\old@startsection=\@startsection
\let\oldl@section=\l@section
\renewcommand{\@startsection}[6]{\old@startsection{#1}{#2}{#3}{#4}{#5}{#6\mathversion{bold}}}
\renewcommand{\l@section}[2]{\oldl@section{\mathversion{bold}#1}{#2}}
\makeatother

\makeatletter
\let\old@makecaption=\@makecaption
\def\@makecaption{\small\old@makecaption}
\makeatother

\def\x{{\tt x}}
\def\y{{\tt y}}
\def\pint{-\!\!\!\!\!\!\!\!\;\int}

\begin{document}

\thispagestyle{empty}
\begin{flushright}\footnotesize
\texttt{ITEP-TH-22/10}\\
\texttt{LPTENS-10/19}\\
\texttt{UUITP-17/10} \vspace{0.8cm}
\end{flushright}

\renewcommand{\thefootnote}{\fnsymbol{footnote}}
\setcounter{footnote}{0}

\begin{center}
{\Large\textbf{\mathversion{bold} Algebraic Curves \\ for Integrable
String Backgrounds\footnote{Based on the talk at "Gauge Fields.
Yesterday, Today, Tomorrow", Moscow, 19-24.01.2010}}\par}

\vspace{1.5cm}

\textrm{
K.~Zarembo$^{1,2}$\footnote{Also at ITEP, Moscow, Russia}}
\vspace{8mm}

\textit{$^{1}$ CNRS -- Laboratoire de Physique Th\'eorique,
Ecole Normale Sup\'erieure\\
24 rue Lhomond, 75231 Paris, France }\\
\texttt{Konstantin.Zarembo@lpt.ens.fr}\vspace{3mm}

\textit{$^{2}$ Department of Physics and Astronomy, Uppsala University\\
SE-751 08 Uppsala, Sweden}\\
\vspace{3mm}


\par\vspace{1cm}

\textbf{Abstract} \vspace{5mm}

\begin{minipage}{14cm}
Many Ramond-Ramond backgrounds which arise in the AdS/CFT
correspondence are described by integrable sigma-models. The
equations of motion for classical spinning strings in these
backgrounds are exactly solvable by finite-gap integration techniques.
We review the finite-gap integral equations
and algebraic curves for coset sigma-models, and then apply the results to the
$AdS_{d+1}$ backgrounds with $d=4$, $3$, $2$, and $1$.
\end{minipage}

\end{center}
\par\vspace{1cm}
\begin{flushright}{\it Dedicated to Andrei Alexeevich Slavnov\\ on occasion of his 70th birthday}
\end{flushright}

\vspace{0.5cm}


\newpage
\setcounter{page}{1}
\renewcommand{\thefootnote}{\arabic{footnote}}
\setcounter{footnote}{0}

\section{Introduction}

The AdS/CFT correspondence is an exact  equivalence of string theory
on the Anti-de-Sitter (AdS) space and conformal field theory on its
boundary  \cite{Maldacena:1998re,Gubser:1998bc,Witten:1998qj}.  One of the surprising features of the AdS/CFT duality is its relationship to integrable systems and exactly solvable models. The integrability is most clearly visible on the string side of the duality and at the classical level. The equations of motion of the string sigma-model for certain AdS background,  notably for $AdS_5\times S^5$ dual to
$\mathcal{N}=4$ super-Yang-Mills (SYM) theory, admit a Lax representation \cite{Bena:2003wd}, with  well-known consequences such as the existence of an infinite set of  conserved charges \cite{Faddeev:1987ph}.
The classical
string integrability  is a manifestation of (or perhaps the reason
for) the full quantum integrability of some AdS/CFT systems in the
large-$N$/free string limit.

The classical solutions of the string sigma-model describe quantum string states
with sufficiently large quantum numbers.  Such states are dual to
local operators in the dual CFT, typically made of a large number of
constituent fields  \cite{Berenstein:2002jq}. The simplest
example is a pointlike string rotating on a big circle
of $S^5$ in $AdS_5\times S^5$ at the speed of light
\cite{Berenstein:2002jq}, which corresponds to a chiral primary
operator in $\mathcal{N}=4$ SYM with large R-charge. More
general spinning strings describe non-BPS  operators with
large energy, spin and angular momentum
\cite{Gubser:2002tv,Frolov:2003qc}\footnote{The case study of the
most interesting solutions can be found in the reviews
\cite{Tseytlin:2003ii,Plefka:2005bk}.}, such as twist-$2$ operators
with infinite spin \cite{Gubser:2002tv}.

Due to integrability, the equations of motion for spinning strings can be integrated by the
finite-gap integration technique \cite{Novikov:1984id}.
At the end, the problem reduces to
a simple set of linear integral equations, which are solvable in terms of
holomorphic integrals on an algebraic curve \cite{Kazakov:2004qf}.
These equations have a direct quantum counterpart. They can be
regarded as the semiclassical limit of the Bethe ansatz equations
for the quantum spectrum of the AdS/CFT system
\cite{Beisert:2005fw}. As we discuss later, the structure of the
finite-gap equations is largely determined by symmetries. This
structure carries over to the quantum Bethe equations
\cite{Beisert:2005fw}, and to some extent to the Y-system/TBA
equations
\cite{Gromov:2009tv,Bombardelli:2009ns,Gromov:2009bc,Arutyunov:2009ur,Bombardelli:2009xz,Gromov:2009at}, at least
in their semiclassical limit
\cite{Gromov:2009tq,Gromov:2010vb}. Recently the integrability
methods were used in computing scattering amplitudes in
$\mathcal{N}=4$ super-Yang-Mills at strong coupling
\cite{Alday:2009yn,Alday:2009dv,Alday:2010vh}. The starting point again is
the classical equations of motion in the sigma-model, albeit
with different boundary conditions.

We review the  construction of the finite-gap integral equations
(classical Bethe equations) for integrable string backgrounds,
mostly following the treatment of the $AdS_5\times S^5$ case in
\cite{Beisert:2005bm}. We first derive the integral equations for an
arbitrary coset sigma-model, using invariant Lie-algebraic language \cite{Babichenko:2009dk}, and
then specialize the construction to particular backgrounds:
$O(3)$ sigma-model \cite{Beisert:2004ag}, $AdS_5\times S^5$
\cite{Beisert:2005bm}, $AdS_4\times CP^3$ \cite{Gromov:2008bz},
$AdS_3\times S^3$ \cite{Babichenko:2009dk}, $AdS_3\times S^3\times
S^3$ \cite{Babichenko:2009dk}, $AdS_2\times S^2$, $AdS_2\times
S^2\times S^2$, and $AdS_2\times S^3$. The finite-gap equations for
the latter three cases have never been derived, and the results
in secs.~\ref{AdS2*S2}-\ref{AdS2*S3} are original. Earlier reviews of
the finite-gap methods in the AdS/CFT correspondence \cite{Zarembo:2004hp,Gromov:2009}
almost exlusively focus on the
$AdS_5\times S^5$ background.

\section{Strings in AdS and integrability}

\subsection{Coset construction}

The Anti-de-Sitter space $AdS_{d+1}$ is a hypersurface:
\begin{equation}\label{X2}
 X^2_{-1}+X^2_{0}-X^2_1-\ldots -X_d^2=1,
\end{equation}
with  a pseudo-Euclidean metric induced from  $\mathbb{R}^{d,2}$. In addition to the metric, the AdS space inherits from $\mathbb{R}^{d,2}$ the action of the $SO(d,2)$ group. This action is transitive, so $AdS_{d+1}$ is actually a homogeneous space of $SO(d,2)$. The little group of any point  (the invariance subgroup of $SO(d,2)$ that leaves this point intact) is $SO(d,1)$. Indeed, the transformations that leave invariant  $X_{-1}=1$, $X_0=\ldots X_d=0$ are rotations of the $X_0,\ldots ,X_d$ coordinates which form $SO(d,1)$. This equippes $AdS_{d+1}$ with the coset structure:  $AdS_{d+1}=SO(d,2)/SO(d,1)$. The AdS space can thus be abstractly defined as a set of equivalence classes of the right $SO(d,1)$ action on $SO(d,2)$: $AdS_{d+1}=\{g\sim gh|g\in SO(d,2), h\in SO(d,1)\}$.

The coset construction is particularly useful in studying the string sigma-model on $AdS_{d+1}$.
One possibility to define the string action is to start with the sigma-model on $SO(d,2)$ and then gauge the right action of $SO(d,1)$ by a non-dynamical gauge field. The gauge transformations are right multiplications from $SO(d,1)$: $g(x)\rightarrow g(x)h(x)$. The gauge fixing is equivalent to picking one representative in each equivalence class or, equivalently, embedding of $AdS_{d+1}$ in $SO(d,2)$. For instance, one can take $g_{AB}=\eta _{AB}+2X_AX_B$ as one such embedding, where $X_A$ are the  $\mathbb{R}^{d,2}$ coordinates in (\ref{X2}), and $\eta _{AB}$ is the $(--+\ldots +)$ metric. The string worldsheet in $AdS_{d+1}$ then is parameterized by $g(x)=g(X(x))$, where $x^{\mathbf{a}}$, $\mathbf{a}=0,1$ are the worldhseet $\tau $ and $\sigma $.

The action of the sigma-model must be gauge-invariant with respect to the $SO(d,1)$ transformations.  This can be achieved by considering the transformation properties of the current
\begin{equation}\label{adscur}
 J_{\mathbf{a}}=g^{-1}\partial _{\mathbf{a}}g.
\end{equation}
The current belongs to the Lie algebra $\mathfrak{so}(d,2)$, and transforms as a gauge connection: $J_\mathbf{a}\rightarrow h^{-1}J_{\mathbf{a}}h+h^{-1}\partial _{\mathbf{a}}h$, the non-homogeneous part of which lies in the $\mathfrak{so}(d,1)$ subalgebra, since $h\in SO(d,1)$. It thus makes sence to decompose the current into two parts:
\begin{equation}\label{}
 J_\mathbf{a}=J_{\mathbf{a}\,0}+J_{\mathbf{a}\,2},
\end{equation}
where $J_{\mathbf{a}\,0}\in \mathfrak{so}(d,1)$ and $J_{\mathbf{a}\,2}$ belongs to the orthogonal complement, which we denote by $\mathfrak{f}$: $\mathfrak{so}(d,2)=\mathfrak{so}(d,1)\oplus \mathfrak{f}$. In the standard lower-right-corner embedding of $\mathfrak{so}(d,1)\subset \mathfrak{so}(d,2)$, $\mathfrak{f}$ is the first row/first coulumn $(d+1)$-dimensional vector. Now, under the gauge tranformations, the $h^{-1}\partial _{\mathbf{a}}h$ term is absorbed into $J_{\mathbf{a}\,0}$ -- this is the gauge field, while the $J_{\mathbf{a}\,2}$ component of the current transforms as the matter field in the adjoint: $J_{\mathbf{a}\,2}\rightarrow h^{-1}J_{\mathbf{a}\,2}h$, and can be used to construct a gauge-invariant string action:
\begin{equation}\label{sigmaomads}
 S=\frac{\sqrt{\lambda }}{8\pi }\int_{}^{}d^2x\,\sqrt{-h}h^{\mathbf{a}\mathbf{b}}\mathop{\mathrm{tr}}
 J_{\mathbf{a}\,2}J_{\mathbf{b}\,2}.
\end{equation}
The coupling constant  $2\pi /\sqrt{\lambda }$ in front (the inverse radius of AdS in the units of $\alpha '$) is related to the 't~Hooft coupling $\lambda $ of the dual CFT.

For any concrete embedding $g(X)$, the coset construction gives the explicit metric on $AdS_{d+1}$: $2ds^2=\mathop{\mathrm{tr}}J^2_2$. For instance, the often-used Poincar\'e coordinates correspond to the following coset parameterization (in the $1+d+1$ decomposition of $\mathbb{R}^{2,d}$ with the standard lower-right-corner embedding of $\mathfrak{so}(d,1)$):
\begin{equation}\label{}
 g=
\begin{pmatrix}
 \frac{z}{2}\left(1+\frac{1+x^2}{z^2}\right) & \frac{x_\nu }{z} &  \frac{z}{2}\left(1-\frac{1-x^2}{z^2}\right) \\
 \frac{x_\mu }{z} & \eta _{\mu \nu }+\frac{2}{(z+1)^2+x^2}\,\,\frac{x_\mu x_\nu }{z} &
 \frac{zx_\mu }{(z+1)^2+x^2}\left(1-\frac{1-x^2}{z^2}\right) \\
 \frac{z}{2}\left(1-\frac{1-x^2}{z^2}\right) & \frac{zx_\nu }{(z+1)^2+x^2}\left(1-\frac{1-x^2}{z^2}\right) &
 1+\frac{1}{2}\,\,\frac{z^3}{(z+1)^2+x^2}\left(1-\frac{1-x^2}{z^2}\right)^2
\end{pmatrix},
\end{equation}
where $\mu ,\nu =0,\ldots ,d-1$, and $\eta _{\mu \nu }$ is the mostly plus Minkowski metric. After many cancellations, the coset construction gives the usual Poincar\'e metric:
\begin{equation}\label{}
 ds^2=\frac{dz^2+dx^2}{z^2}\,.
\end{equation}

However, the abstract language of the coset construction is much more convenient in many respects. Firstly, it allows one to build the necessary supersymmetric completion of the AdS sigma model \cite{Metsaev:1998it}, and, secondly, the coset construction unrevels the hidden integrability structure of the equations of motion. Besides, no particular parameterization is needed to analyze classical solutions. Indeed, the equation of motion for the action (\ref{sigmaomads}) can be written entirely in terms of currents.
The variation of the action gives the conservation condition\footnote{To be more precise, the true Noether current is gauge-invariant and, in terms of the left current $J_\mathbf{a}$, is nonlocal: $k_{\mathbf{a}}=gJ_{\mathbf{a}\,2}g^{-1}$.}:
\begin{equation}\label{adseqmz4}
 2D_{\mathbf{a}}\left(\sqrt{-h}h^{\mathbf{a}\mathbf{b}}J_{\mathbf{b}\,2}\right)=0,
\end{equation}
where $D_{\mathbf{a}}=\partial _{\mathbf{a}}+[J_{\mathbf{a}\,0},\cdot]$ is the covariant derivative.
The $J_2$ component of the current and the gauge field $J_0$ can be regarded as independent variables if the equation of motion is supplemented with the identity that reflects the flatness of (\ref{adscur}). The flatness condition, projected onto $\mathfrak{so}(d,1)$ and $\mathfrak{f}$, decomposes onto two equations:
\begin{eqnarray}\label{adsflatness}
 D_{\mathbf{a}}J_{\mathbf{b}\,2}-D_{\mathbf{b}}J_{\mathbf{a}\,2}&=&0
 \nonumber \\
 F_{\mathbf{a}\mathbf{b}}+[J_{\mathbf{a}\,2},J_{\mathbf{b}\,2}]
 &=&0,
\end{eqnarray}
where $F_{\mathbf{a}\mathbf{b}}=\partial
_\mathbf{a}J_{\mathbf{b}\,0}-\partial
_\mathbf{b}J_{\mathbf{a}\,0}+[J_{\mathbf{a}\,0},J_{\mathbf{b}\,0}]$.
The equation of motion for the metric imposes the Virasoro constraints:
\begin{equation}\label{Viras}
 h^{\mathbf{a}\mathbf{b}}\mathop{\mathrm{tr}} J_{\pm\,\mathbf{a}\,2}J_{\pm\,\mathbf{b}\,2}=0,
\end{equation}
where the $\pm$ superscripts denote the worldsheet light-cone projections:
\begin{equation}\label{chiralvector}
 J_{\pm\,\mathbf{a}\,2}=\left(\delta ^{\mathbf{b}}_{\mathbf{a}}\pm\frac{1}{\sqrt{-h}}\,h_{\mathbf{a}\mathbf{c}}\varepsilon ^{\mathbf{c}\mathbf{b}}\right)J_{\mathbf{b}\,2}.
\end{equation}
The remarkable property of these equations is their complete integrability, which allows to solve them exactly  for quasiperiodic string motions.

\subsection{Integrability}

The geometric origin of integrability in string theory on $AdS_{d+1}$ is an extra $\mathbb{Z}_2$ symmetry of the AdS metric. The metric is obviously invariant under the reflection of the $\mathbb{R}^{2,d}$ embedding coordinates $X_A\rightarrow -X_A$. The $AdS_{d+1}$ manifold thus is a symmetric space. It is interesting to notice that the $\mathbb{Z}_2$ symmetry is not faithfully realized in the Poincar\'e coordinates, which play so important role in the AdS/CFT correspondence. This is because the Poincar\'e coordinates are not geodesically complete and cover just half of the AdS space. Formally, the $\mathbb{Z}_2$ transformation acts as a reflection $z\rightarrow -z$, but  this is not a symmetry of the Poincar\'e patch, in which $z>0$.

In the coset construction, the $\mathbb{Z}_2$ symmetry acts by changing the sign of the $J_2$ component of the current:
\begin{equation}\label{}
 J_{\mathbf{a}\,0}\rightarrow J_{\mathbf{a}\,0},\qquad  J_{\mathbf{a}\,2}\rightarrow -J_{\mathbf{a}\,2}.
\end{equation}
The action (\ref{sigmaomads}) and the equations of motion (\ref{adseqmz4}), (\ref{adsflatness}) are obviously invariant under this transformation. On the more formal, algebraic level, the $\mathbb{Z}_2$ symmetry can be defined as an automorphism of the Lie algebra $\mathfrak{so}(d,2)$ which preserves the coset decoposition $\mathfrak{so}(d,2)=\mathfrak{so}(d,1)\oplus\mathfrak{f}$. The automorphism acts trivially on $\mathfrak{so}(d,1)$, but changes sign of all elements in $\mathfrak{f}$. The fact that this transformation is consistent with the commutation relations of $\mathfrak{so}(d,2)$ is non-trivial, and is of crucial importance for integrability of the model.  To see that the reflection of $\mathfrak{f}$ is a symmetry of $\mathfrak{so}(d,2)$ one can notice that not only $[\mathfrak{so}(d,1),\mathfrak{so}(d,1)]\subset \mathfrak{so}(d,1)$ and $[\mathfrak{so}(d,1),\mathfrak{f}]\subset \mathfrak{f}$, which is true because $\mathfrak{so}(d,1)$ is a subalgebra of $\mathfrak{so}(d,2)$, but also $[\mathfrak{f},\mathfrak{f}]\subset \mathfrak{so}(d,1)$.
Because of the latter property, the flatness condition neatly decomposes into the two equations  (\ref{adsflatness}) and the commutator term appears only in the second of them.

The equations of motion for any symmetric coset admit a Lax representation \cite{Eichenherr:1979ci}:
\begin{equation}\label{eqlaxads}
 L_\mathbf{a}=J_{\mathbf{a}\,0}+\frac{{\tt x}^2+1}{{\tt x}^2-1}\,J_{\mathbf{a}\,2}
 -\frac{2{\tt x}}{{\tt x}^2-1}\,\,\frac{1}{\sqrt{-h}}\,h_{\mathbf{a}\mathbf{b}}\varepsilon
 ^{\mathbf{b}\mathbf{c}}J_{\mathbf{c}\,2}.
\end{equation}
The spectral parameter $\x$ is an arbitrary complex number $\x\neq
\pm 1$. If the currents satisfy the equations of motion, the Lax connection is flat:
\begin{equation}\label{flatness}
 \partial _\mathbf{a}L_\mathbf{b}-\partial
 _\mathbf{b}L_\mathbf{a}+[L_\mathbf{a},L_\mathbf{b}]=0.
\end{equation}
The converse is also true: if the connection $L_\mathbf{a}$ is flat for any
${\tt x}$, the currents satisify the equations (\ref{adseqmz4}), (\ref{adsflatness}). The Virasoro constraints (\ref{Viras}) do not follow from the Lax representation, but are very natural from the point of view of integrability \cite{Faddeev:1985qu}.

The existence of an infinite set of conserved charges, and thus the complete integrability of the model, follows immediately from the Lax representation. The conserved charges are encoded in the monodromy matrix, the Wilson loop of the Lax connection:
\begin{equation}\label{monodram}
 \mathcal{M}({\tt x})={\rm
 P}\exp\oint_{C_{x_*x_*}}dx^\mathbf{a}L_\mathbf{a}(x;{\tt x}),
\end{equation}
The contour of integration $C_{x_*x_*}$  links
the worldsheet, but is otherwise arbitrary. The canonical choice is the equal time section $x^0=x_*^0$, but
because of the flatness condition
(\ref{flatness}) continuous deformations of the contour do not change the monodromy matrix.
The monodromy matrix is a group element of $SO(d,2)$ and transforms by conjugation under the gauge transformations: $\mathcal{M}\rightarrow h^{-1}(x^*)\mathcal{M}h(x^*)$. The shifts of the base point also change the monodromy matrix by conjugation: $\mathcal{M}\rightarrow U^{-1}\mathcal{M}U$, where $U$ is the
monodromy of the flat connection along a curve connecting $x_*$ and $x_*'$.  The eigenvalues of the monodromy matrix do  not change under conjugations,  and are thus gauge-invariant and time-independent. They can be used to define the conserved charges:
\begin{equation}\label{det}
 \mathcal{T}(\x,{\rm z})=\det\left({\rm z}-\mathcal{M}(\x)\right).
\end{equation}
The Laurent expansion of $\mathcal{T}(\x,{\rm z})$ in $\x$ and ${\rm z}$ at an arbitrary reference point produces and infinite set of integrals of motion\footnote{A distinguished choice is ${\rm z}=\infty $, $\x=0$ or $\x=\infty $ or $\x=\pm 1$. In the latter case the conserved charges are integrals of local densities. The expansion at $\x=0$ and $\x=\infty $ starts with the usual Noether charges of the sigma-model.}. The equation (\ref{det}) defines an algebraic curve, generically of an infinite genus, which is the central object in the finite-gap integration method, to be discussed in section~\ref{finitegapsec}.

\subsection{Supersymmetry}

The consistent $AdS_{d+1}$ backgrounds are supersymmetric, which
requires coupling the AdS sigma-model to fermions. In addition,
critical backgrounds of the superstring theory contain  extra
compact factors $M_{9-d}$, and typically are supported by
Ramond-Ramond (RR) fluxes which counter the curvature of
$AdS_{d+1}\times M_{9-d}$. The standard CFT methods of the NSR
formalism are not suitable for the RR backgrounds, and one has to
resort to the Green-Schwarz formalism \cite{Green:1983wt}. The
Green-Schwarz action on $AdS_5\times S^5$ was constructed by Metsaev
and Tseytlin \cite{Metsaev:1998it} with the help of the coset
consrtuction (see \cite{Arutyunov:2009}  for a comprehensive review
of string theory on $AdS_5\times S^5$). The sigma-model is the coset
 $PSU(2,2|4)/SO(4,1)\times SO(5)$ of $PSU(2,2|4)$, the superconformal group of the dual $\mathcal{N}=4$, $D=4$ SYM theory. In addition to the usual metric coupling $G_{MN}\partial X^N\partial X^N$, the Green-Schwarz action should contain
 a fermionic Wess-Zumino term. The coset construction of  $PSU(2,2|4)/SO(4,1)\times SO(5)$ provides a natural candidate because of the $\mathbb{Z}_4$ symmetry
  \cite{Berkovits:1999zq}\footnote{Manifestly $\mathbb{Z}_4$-symmetric formulation of type IIB string theory on $AdS_5\times S^5$ is given in \cite{Roiban:2000yy}.}, which extends  the geometric $\mathbb{Z}_2$ symmetry of the $AdS_5$ manifold.  The integrability of the bosonic string on $AdS_{d+1}$ was a consequence of the $\mathbb{Z}_2$ symmetry, likewise the integrability of the superstring on $AdS_5\times S^5$ is a consequence of he $\mathbb{Z}_4$ symmetry of the  $PSU(2,2|4)/SO(4,1)\times SO(5)$  coset.

 There exists a number of other $\mathbb{Z}_4$ cosets, all of which are integrable and some contain $AdS_{d+1}$ as part of their supergeometry  \cite{Polyakov:2004br,Adam:2007ws,Zarembo:2010sg}.
In the mathematics literature the $\mathbb{Z}_4$ symmetric cosets are called semisymmetric superspaces,  full classification of which was given by Serganova \cite{Serganova}), and can be used as a starting point for a systematic search for consistent integrable string backgrounds  \cite{Zarembo:2010sg}. Many such backgrounds are of the form $AdS_{d+1}\times M$.  This is true for $d=4$ \cite{Metsaev:1998it}, $d=3$ \cite{Arutyunov:2008if,Stefanski:2008ik}, $d=2$ \cite{Rahmfeld:1998zn,Park:1998un,Metsaev:2000mv,Chen:2005uj,Adam:2007ws,Babichenko:2009dk},  and $d=1$ \cite{Zhou:1999sm,Berkovits:1999zq,Verlinde:2004gt,Adam:2007ws}. It thus makes sence to exploit the consequences of integrability in the genral framework of $\mathbb{Z}_4$ cosets, and then specify to the particular cases which are consistent as string backgrounds. Below we review the structure of the $\mathbb{Z}_4$ cosets, the construction of the sigma-model action, its equations of motion, their Lax representation, and the derivation of the finite-gap equations.

A coset $G/H_0$ of the supergroup $G$ possesses a $\mathbb{Z}_4$ symmetry if $\mathfrak{h}_0$ (the Lie algebra of the stabilizer subgroup $H_0$) is invariant under a linear automorphism $\Omega $ of order $4$ that acts on $\mathfrak{g}$, the Lie algebra of the supergroup $G$. The automorphism $\Omega $ is a linear map from $\mathfrak{g}$ to $\mathfrak{g}$ that preserves the Lie bracket. The diagonalization of the $\mathbb{Z}_4$ charge,
\begin{equation}\label{z4action}
 \Omega (\mathfrak{h}_n)=i^n\mathfrak{h}_n,
\end{equation}
defines a $\mathbbm{Z}_4$ decomposition of $\mathfrak{g}$:
\begin{equation}
 \mathfrak{g}=\mathfrak{h}_0\oplus\mathfrak{h}_1\oplus\mathfrak{h}_2\oplus\mathfrak{h}_3.
\end{equation}
Since $\Omega $ preserves the Lie bracket, this decomosition is consistent with the (anti-)com\-mutation relations:
$$[\mathfrak{h}_n,\mathfrak{h}_m\}\subset
\mathfrak{h}_{(n+m)\!\!\!\!\!\mod\! 4}.$$
We also assume that the $\mathbb{Z}_4$ decomposition is consistent with the Grassmann parity, which essentially means that $\Omega ^2=(-1)^F$. Then $\mathfrak{h}_0\oplus\mathfrak{h}_2$ is the bosonic
subalgebra of $\mathfrak{g}$, and $\mathfrak{h}_1$, $\mathfrak{h}_3$
consist of the Grassmann-odd generators.

The embedding of the string worldsheet into $G/H$ is parameterized by a coset
representative $g(x)\in G$, subject to gauge transformations
$g(x)\rightarrow g(x)h(x)$ with $h(x)\in H_0$. The global $G$-valued
transformations act on $g(x)$ from the left: $g(x)\rightarrow
g'g(x)$. The decomposition of the left-invariant current now contains four terms:
\begin{equation}\label{cur}
 J_{\bf a} =g^{-1}\partial _{\bf a} g=J_{{\bf a}\,0 }+J_{{\bf a}\,1 }+J_{{\bf a}\,2
 }+J_{{\bf a}\,3 },
\end{equation}
and there are more freedom in constructing the action. The  possible terms are $J_2J_2$ and $J_1J_3$ contracted with either $h^{\mathbf{ab}}$ or $\varepsilon ^{\mathbf{ab}}$ \footnote{We use the $(+-)$ conventions
for the worldsheet metric, but mostly-plus conventions for the metric in the
target-space. The $\varepsilon $-tensor is defined such that
$\varepsilon ^{01}=1$.}. In the Green-Schwarz-type action, the metric couples to the bosonic currents and the fermionic currents are contracted with $\varepsilon ^{\mathbf{ab}}$:
\begin{equation}\label{action}
 S=\frac{\sqrt{\lambda }}{8\pi }\int_{}^{}d^2x\,\mathop{\mathrm{Str}}
 \left(\sqrt{-h}h^{{\bf a} {\bf b} }J_{{\bf a}\,2 }J_{{\bf b} \,2}+\varepsilon ^{{\bf a} {\bf b} }
 J_{{\bf a}\,1 }J_{{\bf b}\,3 }\right),
\end{equation}
Here $\mathop{\mathrm{Str}}(\cdot \,\cdot )$ is the unique  $G$-invariant bilinear form on $\mathfrak{g}$, which is also invariant under the $\mathbb{Z}_4$ automorphism. The
action is obviously gauge-invariant and $\mathbbm{Z}_4$-symmetric.

The equations of motion and the Maurer-Cartan equations (the flatness condition for the current) form a closed system of seven equations:
\begin{eqnarray}\label{eqmz4}
 2D_\mathbf{a}\left(\sqrt{-h}h^{\mathbf{a}\mathbf{b}}J_{\mathbf{b}\,2}\right)
 -\varepsilon ^{\mathbf{a}\mathbf{b}}[J_{\mathbf{a}\,1},J_{\mathbf{b}\,1}]
 +\varepsilon ^{\mathbf{a}\mathbf{b}}[J_{\mathbf{a}\,3},J_{\mathbf{b}\,3}]
 &=&0
 \nonumber \\
 \left(\sqrt{-h}h^{\mathbf{a}\mathbf{b}}+\varepsilon
 ^{\mathbf{a}\mathbf{b}}\right)
 [J_{\mathbf{a}\,2},J_{\mathbf{b}\,1}]&=&0
 \nonumber \\
 \left(\sqrt{-h}h^{\mathbf{a}\mathbf{b}}-\varepsilon
 ^{\mathbf{a}\mathbf{b}}\right)
 [J_{\mathbf{a}\,2},J_{\mathbf{b}\,3}]&=&0
 \nonumber \\
 \varepsilon ^{\mathbf{a}\mathbf{b}}\left(
 2D_\mathbf{a}J_{\mathbf{b}\,2}
 +[J_{\mathbf{a}\,1},J_{\mathbf{b}\,1}]
 +[J_{\mathbf{a}\,3},J_{\mathbf{b}\,3}]
 \right)&=&0
 \nonumber \\
 \varepsilon ^{\mathbf{a}\mathbf{b}}\left(D_\mathbf{a}J_{\mathbf{b}\,1}
 +[J_{\mathbf{a}\,2},J_{\mathbf{b}\,3}]\right)&=&0
 \nonumber \\
 \varepsilon^{\mathbf{a}\mathbf{b}}\left(D_\mathbf{a}J_{\mathbf{b}\,3}
 +[J_{\mathbf{a}\,2},J_{\mathbf{b}\,1}]\right)&=&0
 \nonumber \\
 F_{\mathbf{a}\mathbf{b}}+[J_{\mathbf{a}\,2},J_{\mathbf{b}\,2}]
 +[J_{\mathbf{a}\,1},J_{\mathbf{b}\,3}]+[J_{\mathbf{a}\,3},J_{\mathbf{b}\,1}]
 &=&0.
\end{eqnarray}
Since the 2d metric only enters the bosonic part of the Lagrangian, the Virasoro constraints (\ref{Viras}) are the same as in the bosonic case.

These equations, as their bosonic cousins, admit a Lax representation \cite{Bena:2003wd}. In order to include fermions one needs to add two extra terms to the Lax connection (\ref{eqlaxads}):
\begin{equation}\label{lax}
 L_\mathbf{a}=J_{\mathbf{a}\,0}+\frac{{\tt x}^2+1}{{\tt x}^2-1}\,J_{\mathbf{a}\,2}
 -\frac{2{\tt x}}{{\tt x}^2-1}\,\,\frac{1}{\sqrt{-h}}\,h_{\mathbf{a}\mathbf{b}}\varepsilon
 ^{\mathbf{b}\mathbf{c}}J_{\mathbf{c}\,2}+\sqrt{\frac{{\tt x}+1}{{\tt x}-1}}\,J_{\mathbf{a}\,1}
 +\sqrt{\frac{{\tt x}-1}{{\tt x}+1}}
 J_{\mathbf{a}\,3}.
\end{equation}
The equations of motion and the Maurer-Cartan equations, will then follow from the flatness condition for $L_{\mathbf{a}}$\footnote{This construction can be generalized to $\mathbb{Z}_n$ symmetric cosets with arbitrary $n$ \cite{Young:2005jv}.}.

\subsection{Finite-gap integration}\label{finitegapsec}

The finite-gap integration method exploits the  Lax representation
of the equations of motion. Instead of dealing with complicated
non-linear PDEs one can study a linear differential equation for the
section of the Lax connection:
\begin{equation}\label{auxlin}
 \left(\frac{d}{dx^1}+L_1(x^0,x^1;\x)\right)\Psi =0.
\end{equation}
The potential in the linear problem, and consequently the currents
can be reconstructed from the wave function $\Psi $
\cite{Novikov:1984id,Faddeev:1987ph}, if necessary, but to calculate
the conserved charges it is enough to understand the spectral
properties of the Dirac-like equation (\ref{auxlin}). The potential
in the Dirac equation is periodic, since the string embedding
coordinates and consequently all the currents are periodic in $x^1$
with the period $2\pi $. The spectrum thus has a band structure with
a series of alternating forbidden and allowed zones in the complex
plane\footnote{The precise locus of the spectrum is determined by
the Hermiticity properties of the Lax connection. To guarantee that
the spectrum lies on the real line the Lax connection should be
self-conjugate in a certain sense. In general the Lax connection
will not be self-conjugate, and we will not assume that the spectrum
is real.} of the spectral parameter $\x$.  The wavefunction is
quasi-periodic in $x^1$, and its monodromy is given by
\begin{equation}\label{}
 \Psi (x^1+2\pi ;\x)=\mathcal{M}(\x)\Psi (x^1;\x),
\end{equation}
where $\mathcal{M}(\x)$ is the monodromy matrix (\ref{monodram}). As
usual, the band structure is determined by the quasi-momenta
$p_l(\x)$, which characterize the periodicity of the wavefunction,
and  are more or less the eigenvalues of $\mathcal{M}(\x)$ (the
precise definition is given below).

The complete analytic characterization of the quasi-momenta for
generic solutions of the sigma-model is sufficient to reconstruct
the spectrum of the solutions themselves. The finite-gap integration
method essentially performs the separation of variables, always
possible in an integrable system. The action variables (the
integrals of motion) are simply expressed in terms of the
quasi-momenta once their analytic structure is understood. The
quasi-momenta, in their turn, are determined by a system of integral
equations which encode the whole semiclassical string spectrum. The
integral equations  were first derived for simple subsectors of
string theory on $AdS_5\times S^5$
\cite{Kazakov:2004qf,Kazakov:2004nh,Beisert:2004ag,SchaferNameki:2004ik,Dorey:2006zj,Dorey:2006mx}
(see \cite{Zarembo:2004hp} for a review), then for the full
semiclassical spectrum \cite{Beisert:2005bm}, and then for strings
on $AdS_4\times CP^3$ \cite{Gromov:2008bz}, $AdS_3\times S^3$
\cite{Babichenko:2009dk,David:2010yg} and $AdS_3\times S^3\times
S^3$ \cite{Babichenko:2009dk}. The finite-gap integral equations can
be solved interms of an algebraic curve. One can also compute
one-loop quantum corrections to arbitrary classical solution using
finite-gap techniques \cite{Gromov:2007ky,Gromov:2008ec} (see
\cite{Gromov:2009} for a review of the algebraic curve technique for
$AdS_5\times S^5$ and of its relationship to quantum Bethe ansatz).
Below we present the derivation of integral equations for any
$\mathbb{Z}_4$ coset.

The monodromy matrix is a group element of $G_{\mathbb{C}}$ which changes by conjugation under gauge transformations and time translations, so its conjugacy class in $G$ is time-independent and gauge-invariant.  The set of
conjugacy classes is isomorphic to the maximal torus of $G$ modulo
Weyl group, and can be conveniently parameterized by choosing a Cartan basis $H_l$, and locally bringing the monodromy matrix to the "diagonal" form:
$$
 \mathcal{M}({\tt x})=U^{-1}({\tt x})\exp\left(p_l({\tt x})H_l
 \right)U({\tt x}).
$$
The quasi-momenta $p_l({\tt x})$ are the gauge-invariant generating
functions for the integrals of motion. They  are defined up to
transformations from the Weyl group and shifts by integer multiples
of $2\pi $, which makes them multiple-valued functions of the
spectral parameter.

The infinite set of integrals of motion constructed from the monodromy matrix contains the ordinary Noether charges generated by the left group multiplication. Indeed, at large spectral parameter, the monodromy matrix expands as
\begin{equation}\label{}
 L_\mathbf{a}=g^{-1}\left(\partial _\mathbf{a}+\frac{1}{\tt x}\,\,
 \frac{2}{h}\,\varepsilon _{\mathbf{a}\mathbf{b}}k^\mathbf{b}\right)g+\ldots
 \qquad ({\tt x}\rightarrow \infty ),
\end{equation}
where
\begin{equation}\label{globch}
 k^\mathbf{a}=g\left(\sqrt{-h}h^{\mathbf{a}\mathbf{b}}J_{\mathbf{b}\,2}
 -\frac{1}{2}\,\varepsilon ^{\mathbf{a}\mathbf{b}}J_{\mathbf{b}\,1}
 +\frac{1}{2}\,\varepsilon
 ^{\mathbf{a}\mathbf{b}}J_{\mathbf{b}\,3}\right)g^{-1}
\end{equation}
are the conserved Noether currents:
\begin{equation}
 \partial _\mathbf{a}k^\mathbf{a}=0.
\end{equation}
The first coefficients of the Laurent expansion of the quasi-momenta
at infinity are thus the Noether charges of the global $G$ symmetry:
\begin{equation}\label{}
 p_l({\tt x})=-\frac{2}{\tt x}\,Q_l \qquad ({\tt x}\rightarrow \infty ).
\end{equation}
Further coefficients of the Leurent expansion constitute an infinite
set of (non-local) integrals of motion responsible for integrability
of the model.  Alternatively, the Laurent expansion at $\x=\pm 1$
generates an infinite set of conserved charges which are integrals
of local densities. Using quasi-momenta, one can also build the
canonical set of action-angle variables
\cite{Dorey:2006mx,Vicedo:2008jy}.

The monodromy matrix $\mathcal{M}({\tt x})$ is a meromorphic
function of the spectral parameter ${\tt x}$ whose only possible
singularities are located at ${\tt x}=\pm 1$. The nature of these
singularities will be discussed later. On the contrary, the analytic structure of quasi-momenta is more complicated. The quasi-momenta are multivalued by their very definition and the diagonalization of the monodromy matrix to a particular Cartan basis may produce branch point with the monodromy in the Weyl group. The ambiguities in diagonalization arise at the endpoints of the forbidden zones of the auxiliary linear problem.  For simplicity, we only
consider the case when the monodromies are elementary Weyl
reflections (including generalized Weyl reflections specific to
supergroups \cite{Dobrev:1985qz,PenkovSerganova}). The solutions with simple monodromies describe elementry string excitations. Any element of the Weyl group can be represented as a product of Weyl reflections and accordingly the solutions with composite monodromies correspond to composite quantum states. Such states arise as solutions of nested Bethe ansatz equations known as stacks
\cite{Beisert:2005di,Gromov:2007ky}.

The Weyl reflection with respect to the $l$th root of the Lie
superalgebra $\mathfrak{g}$ acts on the $l$th quasi-momentum as $
p_l({\tt x})\rightarrow p_l({\tt x})-A_{lm}p_m({\tt x})$, where
$A_{lm}$ is the Cartan matrix of $\mathfrak{g}$. As one encircles a
branch point in the  complex plane of the spectral parameter the
quasi-momentum  changes as
\begin{equation}\label{monodr}
 p_l({\tt x})\rightarrow p_l({\tt
x})-A_{lm}p_m({\tt x})+2\pi n_{l,i}.
\end{equation}
The nature of the branch point  depends on whether the $l$th root of
the superalgebra is bosonic or fermionic. By a slight abuse of
terminology we will also call the corresponding quasi-momentum
bosonic or fermionic, although the Cartan elements $H_m$ are all
even generators of the Lie superalgebra and the quasi-momenta are
even functions of the string embedding coordinates. It is the femion
parity of the root generators $E^\pm_{l}$ that distinguishes bosonic
roots from fermionic. If the root is fermionic, the diagonal element
of the Cartan matrix vanishes, $A_{ll}=0$, and after encircling the
branch point the quasi-momentum  $p_l$ shifts by a known, locally
analytic  function: $p_l\rightarrow p_l+\ldots $. Consequently, the
branch points of a fermionic quasi-momentum are logarithmic. For the
bosonic root $A_{ll}=2$, and the quasi-momentum in addition changes
sign: $p_l\rightarrow -p_l+\ldots $. This means that the singularity
is a square root branch point.

The quasi-momenta are thus meromorphic functions on the
complex plane with punctures at $\x=\pm 1$ and cuts $C_{l,i}$, at the endpoints of which the quasi-momenta have either logarithmic or square root singularities.  For the bosonic,
square-root cuts, the monodromy condition (\ref{monodr}) is
equivalent to an equation for the continuous part of the
quasi-momentum:
\begin{equation}\label{babybethe}
 A_{lm}/\!\!\!p_m({\tt x})=2\pi  n_{l,i},\qquad {\tt x}\in C_{l,i},
\end{equation}
where we define:
\begin{equation}\label{}
 /\!\!\!p_l({\tt x})=\frac{1}{2}\left(p_l({\tt x}+i0 )
 +p_l({\tt x}-i0 )\right).
\end{equation}
The same equation holds at the endpoints of the fermionic cuts, in
which case $p_l(\x)$ actually drops out of the equation.

In addition to the branch cuts, $p_l$ has simple poles at ${\tt
x}=\pm 1$, where the Lax connection (\ref{lax}) itself has a
singularity:
\begin{equation}\label{residueofLax}
 L_\mathbf{a}=\frac{J_{\pm\,\mathbf{a}\,2}}{{\tt x}\pm 1}
 +\ldots \qquad (\x\rightarrow \mp 1) .
\end{equation}
The residue is the chiral projection  (\ref{chiralvector}) of the current.
The quasi-momenta consequently have simple poles at $\x=\pm1$:
\begin{equation}\label{resqm}
 p_l({\tt x})=\frac{1}{2}\,\,\frac{\kappa _l\mp 2\pi m_l}{{\tt x}\pm 1}+\ldots
 \qquad  (\x\rightarrow \mp 1).
\end{equation}
Parameterization of the residues at $\x=1$ and $\x=-1$ by their sum
and difference is a matter of convenience.

The information contained in $p_l({\tt x})$ is actually redundant
because of the $\mathbbm{Z}_4$ symmetry. The symmetry acts on the
flat connection according to (\ref{z4action}). It is  not hard to
see that the $\mathbb{Z}_4$ transformation is equivalent to the
inversion of the spectral parameter:
\begin{equation}\label{}
 \Omega (L_\mathbf{a}({\tt x}))=L_\mathbf{a}(1/{\tt x}).
\end{equation}
The $\mathbb{Z}_4$ action on the Lie algebra can be lifted to the group action
with the help of the exponential map. Thus,
\begin{equation}\label{}
 \Omega (\mathcal{M}(\x))=\mathcal{M}(1/\x).
\end{equation}
Likewise, the exponential map defines the action of the
$\mathbb{Z}_4$  automorphism on  the maximal torus of $G$, albeit up
to the Weyl reflections. Given the $\mathbbm{Z}_4$ action on the
Cartan generators:
\begin{equation}\label{defS}
 \Omega (H_l)=H_mS_{ml},
\end{equation}
we can infer the transformation properties of the quasi-momenta
under the inversion in the spectral-parameter plane:
\begin{equation}\label{inversion}
 p_l(1/{\tt x})=S_{lm}p_m({\tt x}).
\end{equation}
In consequence, the knowledge of the quasi-momenta in the {\it physical
region}, $|{\tt x}|>1$, is sufficient to reconstruct them elsewhere in
the complex plane.

A meromorphic function with the properties listed above is completely determined by its discontinuities at the cuts, which we denote by $2\pi i\rho _l({\tt x})$. For
fermionic cuts, the monodromy condition (\ref{monodr}) actually
determines the discontinuity almost completely, except for its value at the endpoints,
so $\rho _l(\x)$ is then concentrated at the fermionic poles. The
quasi-momenta thus admit a spectral representation:
\begin{equation}\label{}
 p_l({\tt x})=-\frac{\kappa _l{\tt x}+2\pi m_l}{{\tt x}^2-1}
 +\int_{C_l}^{}d{\tt y}\,\frac{\rho _l({\tt y})}{{\tt x}-{\tt y}}
 +\int_{1/C_l}^{}d{\tt y}\,\frac{\tilde{\rho }_l({\tt y})}{{\tt x}-{\tt
 y}}\,,
\end{equation}
where $C_l$ denotes the collection of bosonic cuts and fermionic
poles in the physical domain $|{\tt x}|>1$. The contribution of the
mirror cuts has been separated, because  the inversion symmetry
(\ref{inversion}) determines $\tilde{\rho }$ in terms of $\rho $.

Imposing the condition (\ref{inversion}), we find that  $\kappa _l$
must satisfy\footnote{Since $\Omega^2=(-1)^F $ and $H_l$ are
Grassmann-even, the matrix $S_{lm}$ squares to one and has
eigenvalues equal to one or minus one.}
\begin{equation}\label{eigenkappa}
 S_{lk}\kappa _k=-\kappa _l,
\end{equation}
 that\footnote{Typically, $m_l$ are integers that have the meaning of the string winding numbers  \cite{Kazakov:2004nh}.}
\begin{equation}\label{}
 2\pi m_l=\left(\delta _{lk}-S_{lk}\right)\int_{}^{}\frac{d\x}{\x}\,\rho _k(\x),
\end{equation}
and that
\begin{equation}\label{}
 p_l({\tt x})=-\frac{\kappa _l{\tt x}+2\pi m_l}{{\tt x}^2-1}
 +\int_{}^{}d{\tt y}\,
 \frac{\rho _l({\tt y})}{{\tt x}-{\tt y}}
 -S_{lm}\int_{}^{}\frac{d{\tt y}}{{\tt y}^2}\,\,\frac{\rho_m({\tt y}) }{{\tt x}-\frac{1}{\tt
 y}}\,,
\end{equation}
where the integration contours all lie outside the unit circle.

 The
condition (\ref{babybethe}) becomes a set of integral equations for
the densities:
\begin{equation}\label{classbethe}
 A_{lm}-\!\!\!\!\!\!\!\!\;\int_{}^{}d{\tt y}\,\frac{\rho _m({\tt y})}{{\tt x}-{\tt y}}
 -A_{lk}S_{km}\int_{}^{}\frac{d{\tt y}}{{\tt y}^2}\,\,
 \frac{\rho_m({\tt y}) }{{\tt x}-\frac{1}{\tt
 y}}=A_{lk}\,\frac{\kappa _k{\tt x}+2\pi m_k}{{\tt
 x}^2-1}+2\pi n_{l,i},\qquad {\tt x}\in C_{l,i}.
\end{equation}
The equations hold on the collection of bosonic cuts and fermionic
poles in the complex $\x$ plane. The solutions of these equations
describe the spectrum of quasi-periodic spinning string solutions of
the sigma-model. The conserved charges  can be computed by expanding
the quasi-momenta at infinity.

The constants $\kappa _l$, that determine the source terms in the
integral  equations, originated from the residues of the
quasi-momenta at $\x=\pm 1$ and can be computed from the
semiclassical analysis of the auxiliary linear problem for the Lax
operator \cite{Beisert:2004ag,Zarembo:2004hp}. If the string action
contains just the coset and no other fields, the Virasoro conditions
(\ref{Viras}) imply that the residue of the Lax connection in
(\ref{residueofLax}) is null. The residues of the quasi-momenta
should then satisfy
\begin{equation}\label{nullkappa}
 (\kappa _l\pm 2\pi m_l)A_{lk}(\kappa _k\pm 2\pi m_k)=0.
\end{equation}
This condition, along with eq.~(\ref{eigenkappa}), strongly
constraints the possible form of $\kappa _l$'s. In the particular
cases that we consider below the two conditions will determine
$\kappa _l$'s almost uniquely, up to an overall multiplicative
factor.

The classical finite-gap equations have a direct quantum
counterpart, the Bethe equations for the quantum spectrum of the
sigma-model.  Upon quantization, the cuts in the spectral plane
decompose into sets of discrete points, the Bethe roots, which
satisfy algebraic (or more generally, functional) equations. The
integral equations, which here were derived from the classical
equations of motion of the string, approximate the exact Bethe
equations in the particular thermodynamic limit
\cite{Sutherland:1995zz,DharShastry,Beisert:2003xu}.

\section{Classical Bethe equations for various backgrounds}

\subsection{$O(3)$ sigma-model}

Let us begin with the simplest example, the sigma model on
$S^2=SU(2)/U(1)$. The Dynkin diagram of $SU(2)$ consists of one
node, the Cartan matrix is a number: $A=2$, an the reflection
symmetry  changes the sign of the Cartan generator: $S=-1$. The
integral equation (\ref{classbethe}) then takes the
form\footnote{For simplicity we set $m=0$; see
\cite{Gromov:2006dh,Gromov:2007fn} for the discussion of the $m\neq
0$ case.}:
\begin{equation}\label{o3}
 2-\!\!\!\!\!\!\!\!\;\int_{}^{}d{\tt y}\,\rho (\y)\left(
 \frac{1}{\x-\y}+\frac{1}{\y^2}\,\,\frac{1}{\x-\frac{1}{\y}}
 \right)
 =\frac{2\kappa {\tt x}}{{\tt
 x}^2-1}+2\pi n_{i},\qquad {\tt x}\in C_{i}.
\end{equation}

This singular integral equation describes classical  spinning
strings on $S^2\times \mathbb{R}^1_{\rm time}$
\cite{Beisert:2004ag}. The constant $\kappa$ is related to the
target-space energy of the string:
\begin{equation}\label{kappa}
 \kappa =\frac{2\pi E}{\sqrt{\lambda }}\,.
\end{equation}
The argument goes as follows. If we add the time coordinate to the
sigma-model, with the action (in the conformal gauge)
\begin{equation}\label{}
 S_{\rm time}=-\frac{\sqrt{\lambda }}{4\pi }\int_{-\infty }^{+\infty }dx^0
 \int_{-\pi }^{+\pi }dx^1\,\left(\partial _{\mathbf{a}}X^0\right)^2,
\end{equation}
the solutions carrying energy $E$ will have to $X^0=E
x^0/\sqrt{\lambda } $. The only effect on the non-trivial string
motion on $S^2$ is through  the Virasoro constraints (\ref{Viras})
which acquire  the right-hand side equal to $(\partial _\pm
X^0)^2=E^2/\lambda $. The light-cone components of the sigma-model
currents are then normalized to $E/\sqrt{\lambda }$. Integrating
(\ref{residueofLax}) along $x^1$ to get the monodromy matrix, we
find that
 $$
 p(x)=\frac{\pi E}{\sqrt{\lambda }}\,\,\frac{1}{\x\pm 1}+\ldots \qquad (\x\rightarrow \mp 1),
 $$
and from (\ref{resqm}) we conclude that $\kappa $ is related to $E$ by (\ref{kappa}).

The singular integral equation (\ref{o3}) can be solved in full
generality in terms of hyperelliptic integrals
\cite{Kazakov:2004qf}. The associated hyperelliptic curve is
obtained by gluing together two copies of the complex plane along
the cuts $C_i$. The differential of the quasi-momentum, $dp(\x)$ is
holomorphic on this curve, except for the two double poles at
$\x=\pm 1$.

\subsection{Strings on $AdS_5\times S^5$}

Next we consider the algebraic curve for the Metsaev-Tseytlin
sigma-model \cite{Beisert:2005bm}, which is the
$PSU(2,2|4)/SO(4,1)\times SO(5)$ coset. The Lie superalgebra
$\mathfrak{psu}(2,2|4)$ is a particular real form of
$\mathfrak{psl}(4|4)$ whose Cartan-Weyl basis is described in detail
elsewhere \cite{Kac:1977qb,Frappat:1996pb}. Here we shortly remind
the necessary facts which are useful  for the construction of the
algebraic curve.

It is somewhat easier to deal with $\mathfrak{sl}(4|4)$, an algebra
of complex $(4|4)\times (4|4)$ supermatrices with zero supertrace:
\begin{equation}\label{}
 \mathop{\mathrm{Str}}M\equiv\sum_{i}^{}(-1)^{|i|}M_{ii}=0.
\end{equation}
The parity $|i|$ distinguishes bosonic indices from fermionic.  The
Grassmann parity of the matrix elements is related to the parity of
their indices:
\begin{equation}\label{}
 \left(-1\right)^F\circ M_{ij}=\left(-1\right)^{|i|+|j|}M_{ij}
\end{equation}
In the standard basis $|i|=0$ for $i=1,2,3,4$ and $|i|=1$ for
$i=5,6,7,8$, but it will  prove useful to permute rows and columns
such that the order of bosonic and fermionic indices becomes
different. For the standard choice, the diagonal $4\times 4$ blocks
of $M$ are bosonic, and form $\mathfrak{s}(\mathfrak{u}(4)\oplus
\mathfrak{u}(4))=\mathfrak{so}(6)\oplus\mathfrak{so}(6)\oplus\mathfrak{u}(1)$
subalgebra. The off-diagonal blocks are fermionic. The
$\mathfrak{u}(1)$ factor corresponds to the unit matrix, which has
zero suprtrace and thus belongs to $\mathfrak{sl}(4|4)$, but
obviously commutes with anything else. Therefore this
$\mathfrak{u}(1)$ is central and can be factored out. The resulting
factor-algebra is $\mathfrak{psl}(4|4)$.

\begin{figure}[t]
\centerline{\includegraphics[width=10cm]{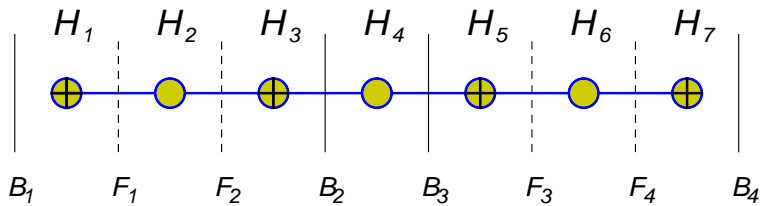}}
\caption{\label{dynka224}\small The Dynkin diagram of
$\mathfrak{psu}(4|4)$ in the preferred basis.}
\end{figure}

The Cartan subalgebra consists of the diagonal supertraceless matrices. The standard choice for the basis of Cartan generators is
\begin{equation}\label{Cartan}
 (H_l)_{ij}=(-1)^{|i|}\delta _{ij}\left(\delta _{il}-\delta _{i\,l+1}\right),\qquad l=1,\ldots ,7,
\end{equation}
with the Cartan matrix
\begin{equation}
 A_{lm}=\mathop{\mathrm{Str}}H_lH_m=\left[(-1)^{|l|}+(-1)^{|l+1|}\right]\delta _{lm}
 -(-1)^{|l|}\delta _{l\,m+1}-(-1)^{|l+1|}\delta _{l\,m-1}.
\end{equation}
This construction is illustrated in
fig.~\ref{dynka224}
(for a non-standard parity assignment, which is distinguished by the
Bethe ansatz equations \cite{Beisert:2005fw}). The eight diagonal
elements of the supermatrix are depicted as  vertical bars ("B" and
"F" stands for the bosonic or fermionic parity). The $l$th node of
the Dynkin diagram connects the $l$th bar with the $(l+1)$st. If the
eigenvalues have the same parity the  node that connects them is
bosonic and the diagonal component $A_{ll}$ of the Cartan matrix is
equal to $2$ or $-2$. If the parity is opposite, the node is
fermionic and $A_{ll}=0$.

The Cartan matrix is not unique for a supergroup
\cite{Kac:1977qb,Frappat:1996pb}. It depends on the parity
assignment for rows an columns, and changes non-trivially under
their permutation.  There is a preferred parity assignment in which
the set of integral equations (\ref{classbethe}) can be directly
related to the Bethe ansatz equations for the quantum spectrum of
the string:
\begin{equation}
 (-1)^{|i|}=(+--++--+).
\end{equation}
This is shown in fig.~\ref{dynka224}. The associated Cartan basis
have the following structure:
\begin{equation}\label{pictcar}
 \begin{tabular}{cccccccc}
 $H_1$ & $H_2$ & $H_3$ & $H_4$ & $H_5$ & $H_6$ & $H_7$ & \\
  \hphantom{-}1 & & & & & & & B\\
  \hline
  \hphantom{-}1 & -1 & & & & & & F\\
  & \hphantom{-}1 & -1 & & & & & F \\
  \hline
  & & -1 & \hphantom{-}1 & & & & B \\
  & & & -1 & \hphantom{-}1 & & & B \\
  \hline
  & & & & \hphantom{-}1 & -1 & & F \\
  & & & & & \hphantom{-}1 & -1 & F \\
  \hline
  & & & & & & -1 & B
 \end{tabular}
\end{equation}
and the Cartan matrix is
\begin{equation}
 A=
\begin{pmatrix}
  & \hphantom{-}1 & & & & &  \\
 \hphantom{-}1 & -2 & \hphantom{-}1 & & & & \\
 & \hphantom{-}1 & & -1 & & & \\
 & & -1 & \hphantom{-}2 & -1 & & \\
 & & & -1 & & \hphantom{-}1 & \\
 & & & & \hphantom{-}1 & -2 & \hphantom{-}1 \\
 & & & & & \hphantom{-}1 &
\end{pmatrix}.
\end{equation}

The $\mathbb{Z}_4$ automorphism,  which defines the coset structure
of $PSU(2,2|4)/SO(4,1)\times SO(5)$,  acts on the supermatrices of
the standard grading $(++++----)$ as \cite{Serganova}:
\begin{equation}
 \Omega \circ
 \begin{pmatrix}
 A & \Theta  \\
 \Psi  & B \\
\end{pmatrix}
=
\begin{pmatrix}
 JA^tJ & -J\Psi  ^tJ\\
 J\Theta  ^t J& JB^tJ \\
\end{pmatrix},\qquad J=
\begin{pmatrix}
 0 & -\mathbbm{1}_{2\times 2} \\
 \mathbbm{1}_{2\times 2} & 0 \\
\end{pmatrix}.
\end{equation}
The action on the diagonal matrices amounts in permutation of their
eigenvalues with simultaneous change of the sign:
\begin{equation}\label{prm}
 B_i\rightarrow -B_{\sigma (i)}, \qquad F_i\rightarrow -F_{\sigma (i)},
\end{equation}
where $\sigma =(3412)$. Taking into account  that the distinguished
basis on fig.~\ref{dynka224} is related to the standard one by
further permutation of indices, we find that in the distinguished
basis  $\sigma =(2143)$. From that, and using the explicit form of
the Cartan generators (\ref{pictcar}), we can infer how the
$\mathbb{Z}_4$ generator acts on the Cartan elements and thus
compute the matrix $S_{lm}$ defined in (\ref{defS}):
\begin{equation}\label{}
 S=
\begin{pmatrix}
  & & \hphantom{-}1 & -1 & & &  \\
  & \hphantom{-}1 & & -1 & & & \\
  \hphantom{-}1 & & & -1 & & & \\
   & & & -1 & & & \\
   & & & -1 & & & \hphantom{-}1 \\
   & & & -1 & & \hphantom{-}1 & \\
   & & & -1 & \hphantom{-}1 & &
\end{pmatrix}.
\end{equation}

Next, we should determine vector $\kappa _l$, which  enters the
right-hand-side of the finite-gap equations (\ref{classbethe}) and
which must satisfy the conditions (\ref{eigenkappa}) and
(\ref{nullkappa}). The eigenvalue equation (\ref{eigenkappa}) has
three linearly independent solutions, one of which is annihilated by
the Cartan matrix $A_{lm}$. Adding to $\kappa _l$  a zero
eigenvector of the Cartan matrix does not change anything, since
$\kappa _l$ enters the equations in the combination $A_{lk}\kappa
_k$, and thus we can concentrate on the orthogonal two-dimensional
subspace. The null condition (\ref{nullkappa}) then has two
solutions, $\kappa _l\propto (1,2,3,4,3,2,1)$ and $\kappa _l\propto
(1,0,-1,0,-1,0,1)$. The second solution appears to be the right one,
and the classical Bethe equations (\ref{classbethe}) take the
following form \cite{Beisert:2005bm}:
\begin{eqnarray}\label{ads5xs5}
2\pi n_{1,i}&=&
\int_{}^{}du(\y)\,\frac{\rho_2 (\y)}{u(\x)-u(\y)}+\int_{}^{}\frac{d\y}{\y^2}\,\,\frac{\rho _4(\y)}{\x-\frac{1}{\y}}
\nonumber \\
2\pi n_{2,i}&=&
\int_{}^{}du(\y)\,\frac{\rho _1(\y)}{u(\x)-u(\y)}-2\pint du(\y)\,\frac{\rho _2(\y)}{u(\x)-u(\y)}
\nonumber \\
&&+\int_{}^{}du(\y)\,\frac{\rho _3(\y)}{u(\x)-u(\y)}
\nonumber \\
2\pi n_{3,i}&=&\int_{}^{}du(\y)\,\frac{\rho _2(\y)}{u(\x)-u(\y)}-\int_{}^{}d\y\,\frac{\rho _4(\y)}{\x-\y}
\nonumber \\
\frac{2\kappa \x}{\x^2-1}+2\pi n_{4,i}&=&
\int_{}^{}\frac{d\y}{\y^2}\,\,\frac{\rho _1(\y)}{\x-\frac{1}{\y}}-\int_{}^{}d\y\,\frac{\rho _3(\y)}{\x-\y}
+2\pint d\y\,\frac{\rho _4(\y)}{\x-\y}-\int_{}^{}d\y\,\frac{\rho _{\bar{3}}(\y)}{\x-\y}
\nonumber \\
&&
+\int_{}^{}\frac{d\y}{\y^2}\,\,\frac{\rho _{\bar{1}}(\y)}{\x-\frac{1}{\y}}
\nonumber \\
2\pi n_{\bar{3},i}&=&\int_{}^{}du(\y)\,\frac{\rho _{\bar{2}}(\y)}{u(\x)-u(\y)}-\int_{}^{}d\y\,\frac{\rho _4(\y)}{\x-\y}
\nonumber \\
2\pi n_{\bar{2},i}&=&\int_{}^{}du(\y)\,\frac{\rho _{\bar{1}}(\y)}{u(\x)-u(\y)}-2\pint du(\y)\,\frac{\rho _{\bar{2}}(\y)}{u(\x)-u(\y)}
\nonumber \\
&&
+\int_{}^{}du(\y)\,\frac{\rho _{\bar{3}}(\y)}{u(\x)-u(\y)}
\nonumber \\
2\pi n_{\bar{1},i}&=&\int_{}^{}du(\y)\,\frac{\rho_{\bar{2}} (\y)}{u(\x)-u(\y)}+\int_{}^{}\frac{d\y}{\y^2}\,\,\frac{\rho _4(\y)}{\x-\frac{1}{\y}}\,.
\end{eqnarray}
Here $u(\x)$ is the Zhukowski variable
\begin{equation}\label{zhuk}
 u(\x)=\x+\frac{1}{\x}\,.
\end{equation}
In the quantum Bethe equations, the Zhukowski variable plays much
more fundamental role than the spectral parameter $\x$. Here it
arises because of the identity
$$
 d\y\left(\frac{1}{\x-\y}-\frac{1}{\y^2}\,\,\frac{1}{\x-\frac{1}{\y}}\right)
 =\frac{du(\y)}{u(\x)-u(\y)}\,.
$$
Once the direct and inversion-symmetry kernels appear with opposite
signs, they combine into the simple Hilbert kernel in terms of the
Zhukowski variable.

\begin{figure}[t]
\centerline{\includegraphics[width=14cm]{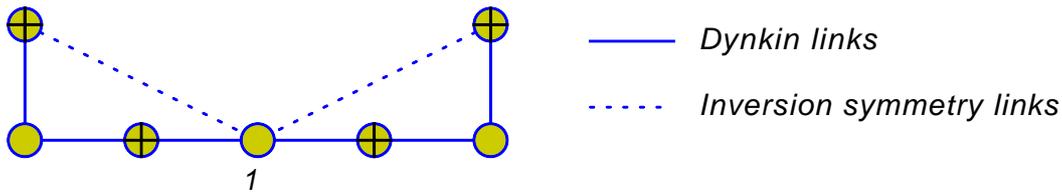}}
\caption{\label{dn224class}\small The Dykin diagram for classical
Bethe equations in $AdS_5\times S^5$.}
\end{figure}

The integral equations are summarized in the diagram in
fig.~\ref{dn224class}.
The couplings between various densities mostly follow the structure
of the Dynkin diagram of $\mathfrak{psu}(2,2|4)$. Addition
interactions arise  due to the inversion symmetry. These
interactions lead to the extra links in fig.~\ref{dn224class} which
connect the central node  to the "wrong" fermionic nodes. The
central node plays rather distinguished role in the classical Bethe
equations (as well as in their quantum counterpart). The source term
appears only in the equation for $\rho _4$. Closely related to this
is the fact that the energy and momentum are computed as moments of
the $\rho _4(\x)$ density:
\begin{eqnarray}\label{}
 \mathcal{E}-\mathcal{J}=\int_{}^{}d\x\,\frac{\rho _4(\x)}{\x^2}
\nonumber \\
\mathcal{P}=\int_{}^{}d\x\,\frac{\rho _4(\x)}{\x}\,.
\end{eqnarray}
Therefore the central node is the only one that  carries energy and
momentum. Roughly speaking, the normalization of $\rho _4$ counts
the total number of excited string oscillator modes in a given
classical solution. The other nodes of the Dynkin diagram are
auxiliary. They determine the flavor structure of the corresponding
string state (in which directions the string oscillates),  but do
not change the energy and momentum.

We will not describe in detail the algebraic curve that solves the
integral equations (\ref{ads5xs5}), see
\cite{Beisert:2005bm,Gromov:2009}. Basically the structure of the
curve follows the Dynkin diagram in fig.~\ref{dynka224}: the curve
is a Riemann surface with eight sheets. Each sheet represents an
eigenvalue of the monodromy matrix and is depicted in
fig.~\ref{dynka224} by a vertical bar. The cuts that connect the
sheets carry the densities and are associated to the nodes of the
Dynkin diagram. We only considered the densities connecting adjacent
sheets, which correspond to the simple roots of the
$\mathfrak{psl}(4|4)$ algebra. The cuts going through several sheets
are stacks combined from several elementary densities
\cite{Beisert:2005di,Gromov:2007ky}.

\subsection{Strings on $AdS_4\times CP^3$}

The type IIA string background  $AdS_4\times CP^3$ is dual to
$\mathcal{N}=6$ Chern-Simons-matter theory in three dimensions
\cite{Aharony:2008ug}, whose superconformal symmetry group is
$OSp(6|4)$. The full Green-Schwarz action on  $AdS_4\times CP^3$ is
rather complicated \cite{Gomis:2008jt}, but upon partially fixing
the kappa-symmetry it reduces to a supercoset sigma-model
\cite{Arutyunov:2008if,Stefanski:2008ik}. The coset,
$OSp(6|4)/U(3)\times SO(3,1)$, has $\mathbb{Z}_4$ symmetry and
therefore is integrable.

The $\mathfrak{osp}(6|4)$ superalgebra can  be represented by
$(6|4)$ supermatrices\footnote{We use a slightly unconventional
definition. Usually $J$ is taken to be $J=i\sigma
^2\otimes\mathbbm{1}$.}:
\begin{equation}\label{whatsosp}
 \mathfrak{osp}(6|4)=\left\{
\begin{pmatrix}
 A & \Theta \\
 \Psi  & B \\
\end{pmatrix}\right|\left.A^t=-A, B^t=JBJ, \Psi =J\Theta  ^t\right\},\qquad J=\mathbbm{1}\otimes i\sigma ^2.
\end{equation}
The Cartan generators can be chosen in the form:
\begin{equation}\label{h}
 H_l=\Lambda _l\otimes i \sigma ^2,\qquad l=1,\ldots ,5,
\end{equation}
where $\Lambda _l$ are diagonal matrices. One should again pick the grading.  The preferred choice for constructing the Bethe equations is $(+--++)$, where the pluses correspond to the $\mathfrak{so}(6)$ and minuses to the $\mathfrak{sp}(4)$ subalgebras. The Cartan genrators are
\begin{equation}\label{gradingosp}
 \begin{tabular}{cccccccc}
 $\Lambda _1$ & $\Lambda _2$ & $\Lambda _3$ & $\Lambda _4$ & $\Lambda _5$  \\
  \hphantom{-}1 & & & & &  B\\
  \hline
  \hphantom{-}1 & -1 & & & & F\\
  & \hphantom{-}1 & -1 & & & F \\
  \hline
  & & -1 & \hphantom{-}1 & \hphantom{-}1&  B \\
  & & & -1 & \hphantom{-}1 & B \\
 \end{tabular}
\end{equation}
and the Cartan matrix ($A_{lm}=\mathop{\mathrm{Str}} H_lH_m$) is
\begin{equation}
 A=
\begin{pmatrix}
  & \hphantom{-}1 & & &   \\
 \hphantom{-}1 & -2 & \hphantom{-}1 & &  \\
 & \hphantom{-}1 & & -1 & -1  \\
 & & -1 & \hphantom{-}2 &   \\
 & & -1 & &  \hphantom{-}2  \\
\end{pmatrix}.
\end{equation}

The $\mathbb{Z}_4$ symmetry acts by conjugation. In the $(+++--)$ grading,
\begin{equation}\label{}
 \Omega \circ
\begin{pmatrix}
 A & \Theta  \\
 \Psi  & B \\
\end{pmatrix}=
\begin{pmatrix}
 -\hat{J}A\hat{J} & -\hat{J}\Theta K \\
 K\Psi \hat{J} & KBK \\
\end{pmatrix},
\end{equation}
where
\begin{equation}
 \hat{J}=
\begin{pmatrix}
  & & & &  &  \hphantom{-}1 \\
  &  & &  &\hphantom{-}1  & \\
  & & &-1& &\\
  & & \hphantom{-}1 & & & \\
  & -1& & & & \\
  -1 & & & & &
\end{pmatrix},\qquad
K=\sigma ^2\otimes \sigma ^1.
\end{equation}
On the matrices of the form (\ref{h}) $\Omega $ acts as
\begin{equation}\label{}
 \Omega \circ \mathop{\mathrm{diag}}\left(\lambda _1,\lambda _2,\lambda _3,\lambda _{\bar{1}},\lambda _{\bar{2}}\right)\otimes i\sigma ^2=\left(-\lambda _3,\lambda _2,-\lambda _1,-\lambda _{\bar{2}},-\lambda _{\bar{1}}\right)\otimes i\sigma ^2.
\end{equation}
Permutating the indices in order to change the grading
to~(\ref{gradingosp}), we find that the $\mathbb{Z}_4$
transformation acts on the eigenvalues of $\Lambda $ as
\begin{equation}\label{}
 \Omega : B_1,F_1,F_2,B_2,B_3\rightarrow -B_2,-F_2,-F_1,-B_1,B_3.
\end{equation}
Given the Cartan generators in~(\ref{gradingosp}), we can now
compute the matrix $S$  from (\ref{defS}):
\begin{equation}\label{invosp}
 S=
\begin{pmatrix}
  & & \hphantom{-}1 & -1 & -1 \\
  & \hphantom{-}1 & & -1  & -1 \\
  \hphantom{-}1 & & & -1 & -1 \\
  & & & & -1 \\
  & & & -1 &
\end{pmatrix}
\end{equation}

\begin{figure}[t]
\centerline{\includegraphics[width=13cm]{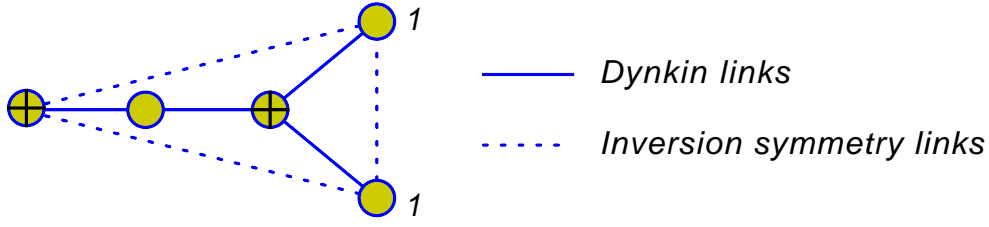}}
\caption{\label{figdynosp}\small The Dynkin diagram for classical Bethe equations in $AdS_4\times CP^3$.}
\end{figure}

Now we can compute $\kappa_l $ with the help of (\ref{eigenkappa})
and (\ref{nullkappa}). The inversion matrix (\ref{invosp}) has two
eigenvectors with eigenvalue $-1$. The null condition
(\ref{nullkappa}), evaluated on the linear combination of these
eigenvectors, has two linearly independent solutions, $\kappa
_l\propto (1,2,3,2,2)$ and $\kappa _l\propto (1,0,-1,0,0)$. The
first solution is unphysical. Substituting the latter solution in
the integral equations (\ref{classbethe}), we get
\cite{Gromov:2008bz}
\begin{eqnarray}\label{}
 2\pi n_{1,i}&=&\int_{}^{}du(\y)\,\frac{\rho _2(\y)}{u(\x)-u(\y)}+\int_{}^{}\frac{d\y}{\y^2}\,\,\frac{\rho _4(\y)}{\x-\frac{1}{\y}}
 +\int_{}^{}\frac{d\y}{\y^2}\,\,\frac{\rho _{\bar{4}}(\y)}{\x-\frac{1}{\y}}
\nonumber \\
 2\pi n_{2,i}&=&\int_{}^{}du(\y)\,\frac{\rho _1(\y)}{u(\x)-u(\y)}-2\pint du(\y)\,\frac{\rho _2(\y)}{u(\x)-u(\y)}
 \nonumber \\
&&+\int_{}^{}du(\y)\,\frac{\rho _3(\y)}{u(\x)-u(\y)}
\nonumber \\
2\pi n_{3,i}&=&\int_{}^{}du(\y)\,\frac{\rho _2(\y)}{u(\x)-u(\y)}-\int_{}^{}{d\y}\,\frac{\rho _4(\y)}{\x-{\y}}
 -\int_{}^{}{d\y}\,\frac{\rho _{\bar{4}}(\y)}{\x-{\y}}
\nonumber \\
\frac{\kappa \x}{\x^2-1}+2\pi n_{4,i}&=&\int_{}^{}\frac{d\y}{\y^2}\,\,\frac{\rho _1(\y)}{\x-\frac{1}{\y}}
-\int_{}^{}d\y\,\frac{\rho _3(\y)}{\x-\y}+2\pint d\y\,\frac{\rho _4(\y)}{\x-\y}
-\int_{}^{}\frac{d\y}{\y^2}\,\,\frac{\rho _4(\y)}{\x-\frac{1}{\y}}
\nonumber \\
&&
+\int_{}^{}\frac{d\y}{\y^2}\,\,\frac{\rho _{\bar{4}}(\y)}{\x-\frac{1}{\y}}
\nonumber \\
\frac{\kappa \x}{\x^2-1}+2\pi n_{\bar{4},i}&=&
\int_{}^{}\frac{d\y}{\y^2}\,\,\frac{\rho _1(\y)}{\x-\frac{1}{\y}}
-\int_{}^{}d\y\,\frac{\rho _3(\y)}{\x-\y}+2\pint d\y\,\frac{\rho _{\bar{4}}(\y)}{\x-\y}
-\int_{}^{}\frac{d\y}{\y^2}\,\,\frac{\rho _{\bar{4}}(\y)}{\x-\frac{1}{\y}}
\nonumber \\
&&
+\int_{}^{}\frac{d\y}{\y^2}\,\,\frac{\rho _{4}(\y)}{\x-\frac{1}{\y}}\,.
\end{eqnarray}
The equations can be summarized by the Dynkin diagram in
fig.~\ref{figdynosp}.
The diagram has two momentum-carrying nodes, which are connected
with each other by an inversion-symmetry link.

\subsection{Strings on $AdS_3\times S^3$}

The Ramond-Ramond backgrounds  of the form $AdS_3\times S^3\times
M_4$ are dual to two-dimensional superconformal field theories. The
extra compact manifold $M_4$ is either a $K3$ surface
\cite{Maldacena:1998re,Berkovits:1999im} ($T^4$ is the simplest
possibility) or $S^3\times S^1$
\cite{Elitzur:1998mm,Cowdall:1998bu,Gauntlett:1998kc,Boonstra:1998yu,Gukov:2004ym}.
These two cases are related, although their superconformal symmetry
is different: $PSU(1,1|2)\times PSU(1,1|2)$ for $M_4=T^4$ or $K3$,
and $D(2,1;\alpha )\times D(2,1;\alpha )$\footnote{The $D(2,1;\alpha
)$ is a one-parameter family of Lie supergroups that continuously
interpolates between $OSp(4|2)$ and $PSU(1,1|2)$
\cite{Kac:1977qb,Frappat:1996pb}.} for $M_4=S^3\times S^1$. We first
consider the simpler case of $M_4=T^4$.

The Green-Schwarz string action on $AdS_3\times S^3$ is a supercoset
sigma-model on $PSU(1,1|2)\times PSU(1,1|2)/SU(1,1)\times SU(2)$
\cite{Rahmfeld:1998zn,Park:1998un,Metsaev:2000mv}, which possesses a
$\mathbb{Z}_4$ symmetry and consequently is integrable
\cite{Chen:2005uj}. In contradistinction to the previously
considered cases, the coset constitutes only part of the geometry
and does not contain the $T^4$ factor. One may wonder if the $T^4$
can at all be coupled to the coset without ruining integrability and
supersymmetry. Fortunately, this is possible, because the
kappa-symmetry of the 10d Green-Schwarz action on $AdS_3\times
S^3\times T^4$ can be fixed in such a way that $T^4$ completely
decouples \cite{Babichenko:2009dk}\footnote{The complete
Green-Schwarz action on $AdS_3\times S^3\times T^4$ can be derived
without the use of the coset construction \cite{Pesando:1998wm}.}.
  At
the classical level one can even treat the coset as a closed sector,
but obviously such a truncation is impossible in the quantum theory,
at least in any direct sense. Incorporating the $T^4$ factor in the
quantum integrability framework remains an unresolved problem.

We  will derive the finite gap equations for the string on
$AdS_3\times S^3$, completely ignoring the $T^4$ factor. The
starting point is the Cartan basis (\ref{Cartan}), which for
$\mathfrak{psu}(1,1|2)$ in the preferred $(-++-)$ grading takes the
form
\begin{equation}\label{carpsu22}
 \begin{tabular}{cccc}
 $H_1$ & $H_2$ & $H_3$ &  \\
  \hphantom{-}1 & & &  F \\
  \hline
  \hphantom{-}1 & -1  & & B \\
  & \hphantom{-}1 & -1 &  B \\
  \hline
  & &  -1 & F
 \end{tabular}
\end{equation}
The Cartan matrix is
\begin{equation}\label{}
 A=
\begin{pmatrix}
 0 & -1 & 0 \\
 -1 & \hphantom{-}2 & -1 \\
 0 & -1 & 0
\end{pmatrix}.
\end{equation}
The denominator of the coset is the direct product $PSU(1,1|2)\times
PSU(1,1|2)$, with the Cartan matrix
\begin{equation}\label{Cartanpsu22}
 A_{\mathfrak{psu}(1,1|2)\oplus\mathfrak{psu}(1,1|2)}=
 A_{\mathfrak{psu}(1,1|2)}\otimes\mathbbm{1}.
\end{equation}

The $\mathbbm{Z}_4$ symmetry generator of  the $PSU(1,1|2)\times
PSU(1,1|2)/SU(1,1)\times SU(2)$ coset is a combination of the
fermion parity and permutation of the two $\mathfrak{psu}(1,1|2)$
factors\footnote{Such automorphism exists for any direct sum of
identical superalgebras. It obviously squares to $(-1)^F$ and
preserves the commutation relations.}
\begin{equation}\label{permferm}
 \Omega =
\begin{pmatrix}
 0 & \mathop{\mathrm{id}} \\
 (-1)^F & 0\\
\end{pmatrix},
\end{equation}
so that the invariant (denominator) subgroup is the bosonic part of
$PSU(1,1|2)$ diagonally embedded into the direct product. On the
Grassmann-even generators the $\mathbb{Z}_4$ symmetry acts just by a
permutation. The $\mathbb{Z}_4$ transformations of the Cartan basis
are then generated by
\begin{equation}\label{Sforperm}
 S=\mathbbm{1}\otimes \sigma ^1.
\end{equation}

The solutions of the eigenvalue  equation (\ref{eigenkappa}) are of
the form  $\kappa \propto v\otimes (1,-1)$. According to
(\ref{nullkappa}), $v$ should satisfy $v^tAv=0$. The null eigenvalue
of $A$, $(1,0,-1)$, can be added with arbitrary coefficient. Up to
this ambiguity which does not affect the resulting source term in
the integral equations, there are two solutions: $v\propto (1,0,1)$
and $v\propto (1,2,1)$. The first solution gives the correct source
term in the Bethe equations.

From these data we find \cite{Babichenko:2009dk} (fig.~\ref{dps}):
\begin{figure}[t]
\centerline{\includegraphics[width=14cm]{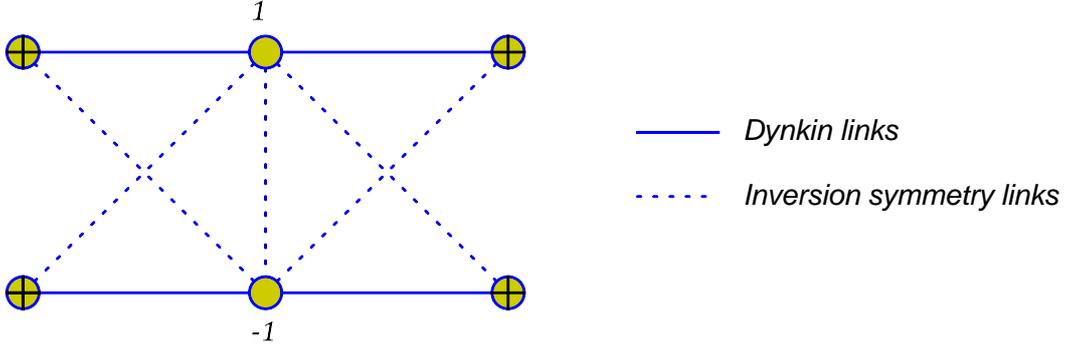}}
\caption{\label{dps}\small The Dynkin diagram for the
$PSU(1,1|2)\times PSU(1,1|2)/SU(1,1)\times SU(2)$ coset.}
\end{figure}
\begin{eqnarray}\label{clbethepsu22_1}
 2\pi n_{1,i}^\pm&=&
 - \int_{}^{}d{\tt y}\,\,\frac{\rho^\pm _2}{{\tt x}-{\tt y}}
 +\int_{}^{}\frac{d{\tt y}}{{\tt y}^2}\,\,\frac{\rho^\mp
 _2}{{\tt x}-\frac{1}{{\tt y}}}
  \nonumber \\
 \pm\frac{2\kappa {\tt x}}
 {{\tt x}^2-1}+2\pi n_{2,i}^\pm
 &=&
 - \int_{}^{}d{\tt y}\,\,\frac{\rho^\pm_1}{{\tt x}-{\tt y}}
 +2-\!\!\!\!\!\!\!\!\,\int_{}^{}d{\tt y}\,\,\frac{\rho^\pm _2}{{\tt x}-{\tt y}}
 - \int_{}^{}d{\tt y}\,\,\frac{\rho^\pm_3}{{\tt x}-{\tt y}}
 \nonumber \\ &&
 +\int_{}^{}\frac{d{\tt y}}{{\tt y}^2}\,\,\frac{\rho^\mp
 _1}{{\tt x}-\frac{1}{{\tt y}}}
 -2\int_{}^{}\frac{d{\tt y}}{{\tt y}^2}\,\,\frac{\rho^\mp
 _2}{{\tt x}-\frac{1}{{\tt y}}}
 +\int_{}^{}\frac{d{\tt y}}{{\tt y}^2}\,\,\frac{\rho^\mp
 _3}{{\tt x}-\frac{1}{{\tt y}}}
 \nonumber \\
 2\pi n_{3,i}^\pm&=&
 - \int_{}^{}d{\tt y}\,\,\frac{\rho^\pm _2}{{\tt x}-{\tt y}}
 +\int_{}^{}\frac{d{\tt y}}{{\tt y}^2}\,\,\frac{\rho^\mp
 _2}{{\tt x}-\frac{1}{{\tt y}}}
\end{eqnarray}
The structure of these equations is very similar to the structure of
the finite-gap equations for $AdS_5\times S^5$ and $AdS_4\times
CP^3$. One may then write down the quantum Bethe equations, of which
(\ref{clbethepsu22_1}) is the classical limit, following the analogy
with the $AdS_5$ and $AdS_4$ cases \cite{Babichenko:2009dk}.
However, here the finite-gap equations describe only a subset of the
degrees of freedom of the string, because the coset constitutes only
a part of the geometry. The string fluctuations in $T^4$ are
describes neither by the finite-gap equations nor by their
conjectured quantum counterpart, which thus capture at best a
subsector of the full string spectrum on $AdS_3\times S^3\times
T^4$.

\subsection{Strings on $AdS_3\times S^3\times S^3$}\label{ads3s3s3sec}

The  $AdS_3\times S^3\times S^3\times S^1$ background is in many
respects similar to $AdS_3\times S^3\times T^4$. The supergroup
$PSU(1,1|2)$ gets replaced by $D(2,1;\alpha )$, but the
$\mathbbm{Z}_4$ symmetry acts as before by eq.~(\ref{permferm}). The
supercoset $D(2,1;\alpha )\times D(2,1;\alpha )/SU(1,1)\times
SU(2)\times SU(2)$ \cite{Babichenko:2009dk} describes the
$AdS_3\times S^3\times S^3$ part of the geometry. The $S^1$ factor
has to be added by hand. One can recover $AdS_3\times S^3\times T^4$
in the limiting case of $\alpha =0$, when one of the three-spheres
blows up to an infinite size and can be recompactified to $T^3$. The
superalgebra $\mathfrak{d}(2,1;\alpha )$ then reduces to
$\mathfrak{psu}(1,1|2)$, up to some Abelian factors. We will not
discuss here the case of arbitrary $\alpha $ (see
\cite{Babichenko:2009dk}) and will concentrate on another special
point, $\alpha =1/2$, when $\mathfrak{d}(2,1;\alpha )$ coincides
with the classical Lie superalgebra $\mathfrak{osp}(4|2)$.

The $\mathfrak{osp}$ Cartan basis (\ref{whatsosp}), (\ref{h}) for
$\mathfrak{osp}(4|2)$ takes the form:
 \begin{equation}\label{42gradingosp}
 \begin{tabular}{cccccc}
  $\Lambda _1$ & $\Lambda _2$ & $\Lambda _3$ & \\
 & -1 & & F  \\ \hline
  \hphantom{-}1& {-}1 & \hphantom{-}1 &  B\\
   {-}1 & & \hphantom{-}1&  B
 \end{tabular}
\end{equation}
The Cartan matrix is
\begin{equation}\label{}
 A=
\begin{pmatrix}
 \hphantom{-}2 & -1 & \hphantom{-}0 \\
 -1 & \hphantom{-}0 & -1 \\
 \hphantom{-}0 & -1 & \hphantom{-}2
\end{pmatrix}.
\end{equation}
As before, the $\mathbb{Z}_4$ generator is given by (\ref{permferm})
and acts on the Cartan generators as in (\ref{Sforperm}).

To find the source term in the integral equations we need to solve
the conditions (\ref{eigenkappa}), (\ref{nullkappa}). There is an
isolated solution $(0,1,0)\otimes (1,-1)$, as well as a
one-parametric family $(a,a^2+(1-a)^2,1-a)\otimes (1,-1)$, which is
spurious and has to be discarded.

\begin{figure}[t]
\centerline{\includegraphics[width=14cm]{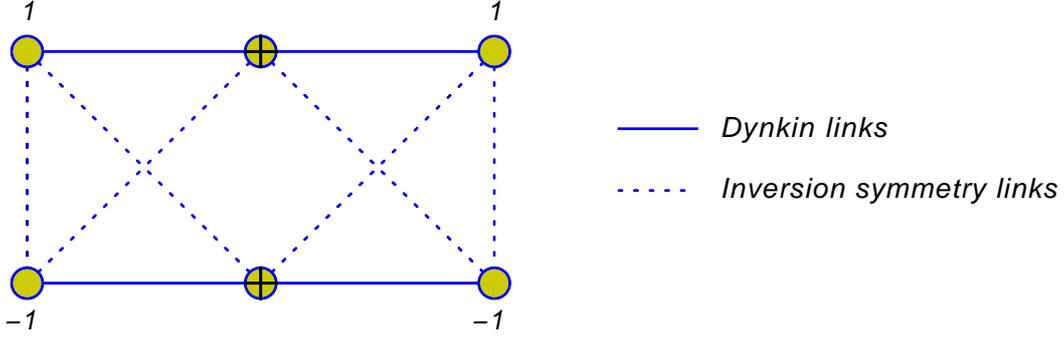}}
\caption{\label{dynkinclass}\small The Dynkin diagram of the
classical Bethe equations.}
\end{figure}

Thus we get for the integral equations:
\begin{eqnarray}\label{clbethe1}
 \pm \frac{2\kappa {\tt x}}
 {{\tt x}^2-1}+2\pi n_{1,i}^\pm&=&
 2 -\!\!\!\!\!\!\!\!\,\int_{}^{}d{\tt y}\,\,\frac{\rho^\pm _1}{{\tt x}-{\tt y}}
 -\int_{}^{}d{\tt y}\,\,\frac{\rho ^\pm_2}{{\tt x} - {\tt y}}
 \nonumber \\
 &&-2\int_{}^{}\frac{d{\tt y}}{{\tt y}^2}\,\,\frac{\rho^\mp
 _1}{{\tt x}-\frac{1}{{\tt y}}}
 + \int_{}^{}\frac{d{\tt y}}{{\tt y}^2}\,\,\frac{\rho ^\mp_2}
 {{\tt x} - \frac{1}{{\tt y}}}
 \nonumber \\
 2\pi n_{2,i}^\pm&=&
 -\int_{}^{}d{\tt y}\,\,\frac{\rho^\pm
 _1}{{\tt x}-{\tt y}}
 -\int_{}^{}d{\tt y}\,\,\frac{\rho ^\pm_3}{{\tt x} - {\tt y}}
 \nonumber \\
 &&+\int_{}^{}\frac{d{\tt y}}{{\tt y}^2}\,\,\frac{\rho^\mp
 _1}{{\tt x}-\frac{1}{{\tt y}}}
 +\int_{}^{}\frac{d{\tt y}}{{\tt y}^2}\,\,\frac{\rho ^\mp_3}
 {{\tt x} - \frac{1}{{\tt y}}}
 \nonumber \\
 \pm \frac{2\kappa {\tt x}}
 {{\tt x}^2-1}+2\pi n_{3,i}^\pm&=&
 2-\!\!\!\!\!\!\!\!\,\int_{}^{}d{\tt y}\,\,\frac{\rho^\pm _3}{{\tt x}-{\tt y}}
 -\int_{}^{}d{\tt y}\,\,\frac{\rho ^\pm_2}{{\tt x} - {\tt y}}
 \nonumber \\
 &&-2\int_{}^{}\frac{d{\tt y}}{{\tt y}^2}\,\,\frac{\rho^\mp
 _3}{{\tt x}-\frac{1}{{\tt y}}}
 +\int_{}^{}\frac{d{\tt y}}{{\tt y}^2}\,\,\frac{\rho ^\mp_2}
 {{\tt x} - \frac{1}{{\tt y}}}\,.
\end{eqnarray}
The Dynkin diagram for these equations is shown in
fig.~\ref{dynkinclass}.
The coset does not include the $S^1$ factor and so string
fluctuations in the $S^1$ directions are not captured by the
integral equations. Moreover, in this case the finite-gap equations
do not describe the massless degrees of freedom which belong to the
coset \cite{Babichenko:2009dk}.

\subsection{Strings on $AdS_2\times S^2$}\label{AdS2*S2}

The Green-Schwarz action on  $AdS_2\times S^2$  is described by the
$PSU(1,1|2)/U(1)\times U(1)$ coset
\cite{Zhou:1999sm,Berkovits:1999zq}. In the supermatrix
representation of $\mathfrak{su}(1,1|2)$, the $\mathbb{Z}_4$
symmetry acts as
\begin{equation}\label{}
 \Omega \circ
\begin{pmatrix}
 A & \Theta \\
 \Psi  & B \\
\end{pmatrix}=
\begin{pmatrix}
 \sigma ^2A \sigma ^2& i\sigma ^2\Theta \sigma ^2 \\
 -i\sigma ^2\Psi \sigma ^2 & \sigma ^2B\sigma ^2 \\
\end{pmatrix},
\end{equation}
from which we can  infer the $\mathbb{Z}_4$ transformation of the
Cartan elements (\ref{carpsu22}):
\begin{equation}\label{}
 S=
\begin{pmatrix}
 & & -1 \\
 &-1 & \\
 -1 & &
\end{pmatrix}.
\end{equation}

\begin{figure}[t]
\centerline{\includegraphics[width=14cm]{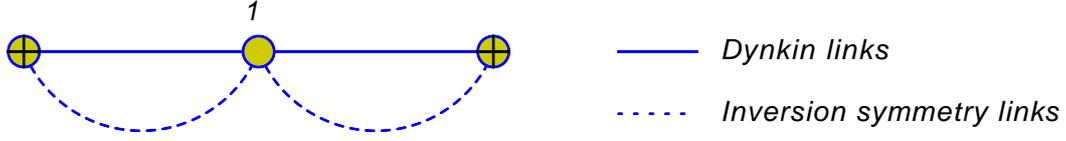}}
\caption{\label{dynads2}\small The Dynkin diagram for $AdS_2\times S^2$.}
\end{figure}

In the two-dimensional space orthogonal  to the zero eigenvalue of
the Cartan matrix $(1,0,-1)$ the conditions (\ref{eigenkappa}) and
(\ref{nullkappa}) have two solutions: $\kappa _l\propto (1,2,1)$ and
$\kappa _l\propto (1,0,1)$. The former leads to momentum-carrying
fermion nodes and has to be discarded. The latter solution gives the
following integral equations:
\begin{eqnarray}\label{}
 2\pi n_{1,i}&=&\int_{}^{}d\y\,\rho_2 (\y)\left(\frac{1}{\x-\y}+\frac{1}{\y^2}\,\,\frac{1}{\x-\frac{1}{\y}}\right)
\nonumber \\
\frac{2\kappa \x}{\x^2-1}+2\pi n_{2,i}&=&
2\pint_{}^{}d\y\,\rho_2 (\y)\left(\frac{1}{\x-\y}+\frac{1}{\y^2}\,\,\frac{1}{\x-\frac{1}{\y}}\right)
\nonumber \\
&&
-\int_{}^{}d\y\,\rho_1 (\y)\left(\frac{1}{\x-\y}+\frac{1}{\y^2}\,\,\frac{1}{\x-\frac{1}{\y}}\right)
\nonumber \\
&&
-\int_{}^{}d\y\,\rho_{\bar{1}} (\y)\left(\frac{1}{\x-\y}+\frac{1}{\y^2}\,\,\frac{1}{\x-\frac{1}{\y}}\right)
\nonumber \\
2\pi n_{\bar{1},i}&=&\int_{}^{}d\y\,\rho_2
(\y)\left(\frac{1}{\x-\y}+\frac{1}{\y^2}\,\,\frac{1}{\x-\frac{1}{\y}}\right).
\end{eqnarray}
The Dynkin diagram for these equations is shown in
fig.~\ref{dynads2}.

When the fermionic densities are switched off and  there are no
stacks, the equation for  $\rho _2$ reduces to the integral equation
for the $O(3)$ sigma-model (\ref{o3}). This is not surprising, since
the $O(3)$ sigma-model describes a subset of string configurations
which move on $S^2$ and sit at the centre of $AdS_2$.

\subsection{$AdS_2\times S^2\times S^2$}

The supersymmetric sigma-model  on $AdS_2\times S^2\times S^2$ is a
$\mathbb{Z}_4$ coset $OSp(4|2)/U(1)\times U(1)\times U(1)$
\cite{Zarembo:2010sg} with the $\mathbb{Z}_4$ generator
\begin{equation}\label{qq}
 \Omega \circ
\begin{pmatrix}
 A & \Theta \\
 \Psi  & B \\
\end{pmatrix}=
\begin{pmatrix}
 QAQ & iQ\Theta \sigma ^3 \\
 -i\sigma ^3\Psi Q & \sigma ^3B\sigma ^3
\end{pmatrix},
\end{equation}
where $Q=\mathop{\mathrm{diag}}(1,-1,1,-1)$. The elements  of the
Cartan subalgebra (\ref{h}), (\ref{42gradingosp}) transform just by
the reflection of sign:
\begin{equation}\label{z4simple}
 S=-1.
\end{equation}

\begin{figure}[t]
\centerline{\includegraphics[width=14cm]{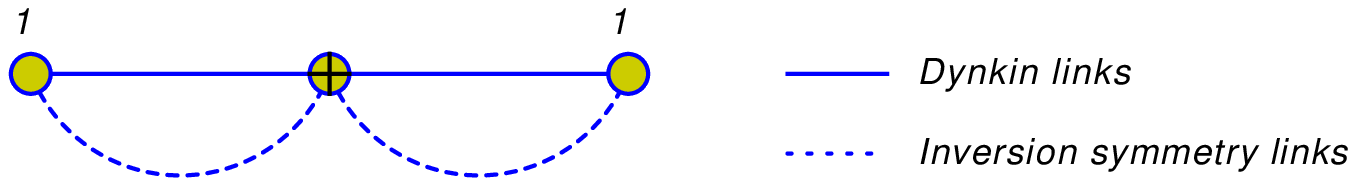}}
\caption{\label{42Dynkin}\small The Dynkin diagram for $AdS_2\times S^2\times S^2$.}
\end{figure}

The solution of the null condition (\ref{nullkappa}) is the same as in sec.~\ref{ads3s3s3sec}: $\kappa _l\propto (0,1,0)$. Combining the Cartan matrix of $\mathfrak{osp}(4|2)$ with the $\mathbb{Z}_{4}$ generator (\ref{z4simple}) and this $\kappa $, we arrive at the following integral equations:
\begin{eqnarray}\label{}
 \frac{\kappa \x}{\x^2-1}+2\pi n_{1,i}&=&
 2\pint_{}^{}d\y\,\rho_1 (\y)\left(\frac{1}{\x-\y}+\frac{1}{\y^2}\,\,\frac{1}{\x-\frac{1}{\y}}\right)
\nonumber \\
&&
 -\int_{}^{}d\y\,\rho_2 (\y)\left(\frac{1}{\x-\y}+\frac{1}{\y^2}\,\,\frac{1}{\x-\frac{1}{\y}}\right)
\nonumber \\
2\pi n_{2,i}&=&
\int_{}^{}d\y\,\rho_1 (\y)\left(\frac{1}{\x-\y}+\frac{1}{\y^2}\,\,\frac{1}{\x-\frac{1}{\y}}\right)
\nonumber \\
&&
+\int_{}^{}d\y\,\rho_{\bar{1}} (\y)\left(\frac{1}{\x-\y}+\frac{1}{\y^2}\,\,\frac{1}{\x-\frac{1}{\y}}\right)
\nonumber \\
\frac{\kappa \x}{\x^2-1}+2\pi n_{\bar{1},i}&=&
 2\pint_{}^{}d\y\,\rho_{\bar{1}} (\y)\left(\frac{1}{\x-\y}+\frac{1}{\y^2}\,\,\frac{1}{\x-\frac{1}{\y}}\right)
\nonumber \\
&&
 -\int_{}^{}d\y\,\rho_2
 (\y)\left(\frac{1}{\x-\y}+\frac{1}{\y^2}\,\,\frac{1}{\x-\frac{1}{\y}}\right).
\end{eqnarray}
The associated Dynkin diagram is shown in fig.~\ref{42Dynkin}.

\subsection{$AdS_2\times S^3$}\label{AdS2*S3}

\begin{figure}[t]
\centerline{\includegraphics[width=14cm]{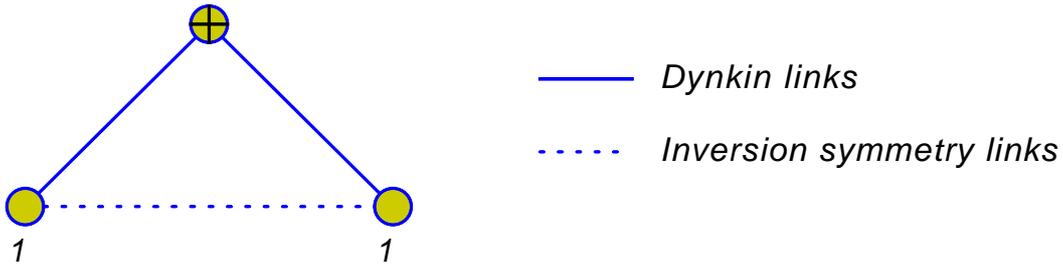}}
\caption{\label{lastdyn}\small The Dynkin diagram for $AdS_2\times S^3$.}
\end{figure}

The $AdS_2\times S^3$ coset is $OSp(4|2)/SO(3)\times U(1)$
\cite{Zarembo:2010sg}, for which the $\mathbb{Z}_4$ generator is
given by (\ref{qq}) with ${Q}=\mathop{\mathrm{diag}}(-1,1,1,1)$. The
$\mathbb{Z}_4$ symmetry then acts on the Cartan generators as
$\Omega \circ\mathop{\mathrm{diag}}(\lambda _1,\lambda _2,\lambda
_3)\otimes i\sigma ^2=\mathop{\mathrm{diag}}(-\lambda _1,-\lambda
_2,\lambda _3)\otimes i\sigma ^2$. For the Cartan basis
(\ref{42gradingosp}), this gives:
\begin{equation}\label{}
 S=
\begin{pmatrix}
 & & -1 \\
 &-1 & \\
 -1 & &
\end{pmatrix}.
\end{equation}
Solving the null condition for the $-1$ eigenvectors of $S$  we find
two solutions: the correct solution $\kappa _l\propto (0,1,0)$, and
the spurious solution $\kappa _l\propto (1,1,1)$ which leads to
momentum-carrying fermion nodes. The integral equations are
\begin{eqnarray}\label{}
\frac{2\kappa \x}{\x^2-1}+2\pi n_{1,i}&=&
 2\pint d\y\,\rho _1(\y)\,\frac{1}{\x-\y}-\int_{}^{}d\y\,\rho _2(\y)\left(\frac{1}{\x-\y}+\frac{1}{\y^2}\,\,\frac{1}{\x-\frac{1}{\y}}\right)
\nonumber \\
&&
 +2\int d\y\,\rho _{\bar{1}}(\y)\,\frac{1}{\y^2}\,\,\frac{1}{\x-\frac{1}{\y}}
\nonumber \\
2\pi n_{2,i}&=&
\int_{}^{}d\y\,\rho _1(\y)\left(\frac{1}{\x-\y}+\frac{1}{\y^2}\,\,\frac{1}{\x-\frac{1}{\y}}\right)
\nonumber \\
&&
+\int_{}^{}d\y\,\rho _{\bar{1}}(\y)\left(\frac{1}{\x-\y}+\frac{1}{\y^2}\,\,\frac{1}{\x-\frac{1}{\y}}\right)
\nonumber \\
\frac{2\kappa \x}{\x^2-1}+2\pi n_{\bar{1},i}&=&
 2\pint d\y\,\rho _{\bar{1}}(\y)\,\frac{1}{\x-\y}-\int_{}^{}d\y\,\rho _2(\y)\left(\frac{1}{\x-\y}+\frac{1}{\y^2}\,\,\frac{1}{\x-\frac{1}{\y}}\right)
\nonumber \\
&&
 +2\int d\y\,\rho _{1}(\y)\,\frac{1}{\y^2}\,\,\frac{1}{\x-\frac{1}{\y}}\,.
\end{eqnarray}
The Dynkin diagram is shown in fig.~\ref{lastdyn}.

\section{Quantization}

The finite-gap integral equations can be regarded as a classical
limit of the asymptotic Bethe equations for the quantum spectrum of
the string. The first-principles derivation of the Bethe equations
requires the knowledge of the exact worldsheet S-matrix
\cite{Staudacher:2004tk} in the light-cone gauge. As the experience
with the $AdS_5$ and $AdS_4$ backgrounds demonstrates the exact
S-matrices in AdS/CFT \cite{Beisert:2005tm,Ahn:2008aa} are almost
uniquely determined by symmetries, unitarity and the crossing
condition \cite{Janik:2006dc}, like in many other integrable models
\cite{Zamolodchikov:1978xm,Dorey:1996gd} (see \cite{Arutyunov:2009}
for a thorough review of the bootstrap program for strings on
$AdS_5\times S^5$). The diagonalization of the S-matrix for the
$n$-particle scattering then yields the Bethe equations. However, a
simple set of mnemonic rules allows one to reconstruct the Bethe
equations directly from their semiclassical limit, the finite-gap
equations of the string sigma-model.

The quantum Bethe equations inherit without change the
Dynkin-diagram structure of the finite-gap equations\footnote{The
relationship between the Bethe equations and Dynkin diagrams for
relativistic systems and spin chains was recognized long ago
\cite{Ogievetsky:1986hu}.}. In the latter, the Dynkin diagram
parameterizes the integral kernels, which are of two types: (i) the
Hilbert kernel $1/(\x-\y)$ on the Dynkin links (the first term in
(\ref{classbethe})) and (ii) the inversion-symmetry kernel
$\y^{-2}/(\x-1/\y)$ (the second term in (\ref{classbethe})).
Sometimes the two combine to form the Hilbert kernel
$1/(u(\x)-u(\y))$ in the Zhukowski variables (\ref{zhuk}).

The Zhukowski variable $u$ plays  a fundamental role in the quantum
Bethe ansatz for $AdS/CFT$, the same as the rapidity plays in the
relativistic integrable models.  In addition to the spectral
parameter $\x(u)$, given by eq.~(\ref{zhuk}), it is convenient to
introduce the shifted variables \cite{Beisert:2005fw}
\begin{equation}\label{}
 \x^\pm+\frac{1}{\x^\pm}=u\pm\frac{i}{2h(\lambda )}\,.
\end{equation}
In the $AdS_5/CFT_4$ correspondence, the function $h(\lambda )$
equals to half of the string tension and is related to the 't~Hooft
coupling $\lambda $ of $\mathcal{N}=4$ SYM by $h(\lambda
)=\sqrt{\lambda }/4\pi $. In other cases $h(\lambda )$ apparently
has a more complicated functional dependence on the string tension
and scales as $\sqrt{\lambda }$ only in the semiclassical limit
$\lambda \rightarrow \infty $. The energy and momentum of a particle
with rapidity $u$ are given by
\begin{equation}\label{enmom}
 \,{\rm e}\,^{\frac{ip}{2h}}=\frac{\x^+}{\x^-}\,\qquad
 \varepsilon =\frac{i}{2}\left(\frac{1}{\x^+}-\frac{1}{\x^-}\right).
\end{equation}
The semiclassical limit is the limit of large $h$.

The Bethe equations are generalized periodicity conditions on the
multiparticle states, which take into account pairwise scattering.
Their generic form is
\begin{equation}\label{qBethe}
 \left(\frac{\x^+(u_{k,l})}{\x^-(u_{k,l})}\right)^{n_lL}=\prod_{(j,m)\neq(k,l)}^{}
 S_{lm}(u_{k,l},u_{j,m}).
\end{equation}
Here $L$ is the length of the string (some large quantum number
related to the angular momentum). The solutions are sets of
rapidities $\{u_{k,l}\}$, from which one can determine the energy
and the momentum of the corresponding quantum state by summing
(\ref{enmom}) over all momentum-carrying rapidities, those for which
$n_l\neq 0$\footnote{The contribution of $u_{k,l}$ to the total
momentum is weighted with the factor $n_l$  and to the total energy
with the factor $n_l^2$.}. The integral equations of the finite-gap
method arise from the thermodynamic limit of the quantum Bethe
equations, namely in the limit when the number of rapidities scales
with $h\rightarrow \infty $ such that they condense on the cuts in
the complex plane and can be characterized by continuous densities
\begin{equation}\label{}
 \rho _l(\x)=\frac{1}{h}\sum_{k}^{}\frac{\x_{k,l}^2}{\x_{k,l}^2-1}\,\delta
 (\x-\x_{k,l}).
\end{equation}
At large $h$ the momentum factor in the left-hand side of the Bethe
equations expands as
\begin{equation}\label{}
 \frac{\x^+}{\x^-}\approx 1+\frac{i}{h}\,\,\frac{\x}{\x^2-1}\,,
\end{equation}
and produces the force term in the finite-gap equations
(\ref{classbethe}).

The scattering factors $S_{lm}(\x,\y)$ are determined from the
Dynkin diagram, namely from the Cartan matrix $A_{lm}$ and from its
product with the inversion-symmetry matrix, $(AS)_{lm}$. The
structures that can appear are of four types:
\begin{itemize}
\item {\bf Bosonic nodes:}
Each bosonic node of the Dynkin diagram ($A_{ll}=\pm 2$) is
associated with a factor\footnote{Taken as is for $A_{ll}=2$ and
inverted for $A_{ll}=-2$.}
\begin{equation}\label{}
 \frac{\x^+-\y^-}{\x^--\y^+}\,\,
 \frac{1-\frac{1}{\x^+\y^-}}{1-\frac{1}{\x^-\y^+}}
 \approx
 1+\frac{2i}{h}\left(\frac{\y^2}{\y^2-1}\,\,\frac{1}{\x-\y}
 -\frac{1}{\y^2-1}\,\,\frac{1}{\x-\frac{1}{\y}}
 \right),
\end{equation}
which can also be written as
\begin{equation}\label{}
 \frac{u-v+\frac{i}{h}}{u-v-\frac{i}{h}}\,,
\end{equation}
where $u$ and $v$ are the Zhukowski variables associated with $\x$
and $\y$. As one can see from its large-$h$ expansion, this factor
contributes $2$ to the Hilbert kernel associated with $A_{ll}$, as
it should, but also $-2$ to the inversion kernel associated with
$(AS)_{ll}$. When $(AS)_{ll}=A_{ll}$, the two terms in the
finite-gap equations combine into a single Zhukowski-type kernel.
This is not always the case, and if there is a mismatch, an extra
factor should be added to the Bethe equations. This factor is the
BES dressing phase \cite{Beisert:2006ez,Beisert:2006ib} $\sigma_{\rm
BES}(u,v)$:
\begin{equation}\label{}
 \sigma _{\rm BES}(u,v)=\exp\left(-i\sum_{r,s=\pm}^{}rs\,\chi (\x^r,\y^s)\right).
\end{equation}
The BES phase is a fairly complicated function of rapidities, which
admits the following integral representation \cite{Dorey:2007xn}:
\begin{equation}\label{}
 \chi (\x,\y)=\frac{i}{4\pi ^2}\oint_{|z|=1,|w|=1}
 \frac{dz\,dw}{(\x-z)(\y-w)}\,\,\ln
 \frac{\Gamma\left(1+ih\left(z+\frac{1}{z}-w-\frac{1}{w}\right)\right)}
 {\Gamma\left(1-ih\left(z+\frac{1}{z}-w-\frac{1}{w}\right)\right)}\,.
\end{equation}
The BES phase simplifies in the large-$h$ limit when it reduces to
the AFS phase \cite{Arutyunov:2004vx}:
\begin{equation}\label{}
 \chi
 (\x,\y)\approx
 h(\x-\y)\left(1-\frac{1}{\x\y}\right)\ln\left(1-\frac{1}{\x\y}\right).
\end{equation}
When $h$ is large, 
\begin{equation}
 \x^\pm=\x\pm\frac{i}{2h}\,\,\frac{\x^2}{\x^2-1}+O\left(\frac{1}{h^2}\right),
\end{equation}
and
\begin{equation}
 \sum_{sr=\pm}^{}sr\,\chi (\x^s,\x^r)=-\frac{1}{h^2}\,\,\frac{\x^2}{\x^2-1}\,\,
 \frac{\y^2}{\y^2-1}\,\,\frac{\partial ^2\chi }{\partial \x\,\partial \y}
+O\left(\frac{1}{h^4}\right).
\end{equation}

which gives
\begin{equation}\label{}
 \sigma _{\rm BES}(\x,\y)\approx 1+\frac{i}{h}
 \left(\frac{1}{\y^2-1}\,\,\frac{1}{\x-\frac{1}{\y}}
 -\frac{1}{\y^2-1}\,\,\frac{\x}{\x^2-1}
 \right).
\end{equation}
The second term in the brackets, upon integration over $\y$, shifts
the coefficient of the source term in the classical Bethe equations.
The first term compensates for the mismatch in the coefficient of
the Hilbert and inversion kernels, and thus the BES factor must be
raised to the power $A_{ll}-(AS)_{ll}$. For the fermionic nodes,
there is no self-scattering term since $A_{ll}=0$, and in all the
cases that we have encountered also $(AS)_{ll}=0$.
\item {\bf Bosonic inversion links:}
If two different bosonic nodes are connected by the inversion link,
we associate to it a BES phase factor $\sigma _{\rm
BES}^{-(AS)_{lm}}(u_{k,l},u_{j,m})$. In all the cases when this
happens, both nodes are momentum-carrying. This is important for
consistency, because the BES phase in the semiclassical limit
contributes to the source term in the finite-gap equations, which
normally originates from the momentum factor in the Bethe equations.
Taking into account the contribution of the BES phase, we can deduce the relationship between the parameters $\kappa _l$, that enter the finite-gap equations (\ref{classbethe}), and the length of the string, that enters the quantum Bethe equations (\ref{qBethe}):
\begin{equation}
 A_{lm}\kappa _m=\frac{n_lL}{h}+A_{lk}\left(\delta _{km}-S_{km}\right)
 \int_{}^{}d\x\,\,\frac{\rho _m(\x)}{\x^2}\,.
\end{equation}
Another special property of the Dynkin diagrams that come out of the
finite-gap integration procedure is that bosonic and fermionic nodes
alternate, such that the same-parity nodes are  never connected by
normal Dynkin links.
\item{\bf Dynkin boson-fermion links:}
A normal Dynkin link between a bosonic and fermionic node is
associated with a factor
\begin{equation}\label{}
 \frac{\x-\y^+}{\x-\y^-}\approx
 1-\frac{i}{h}\,\,\frac{\y^2}{\y^2-1}\,\,\frac{1}{\x-\y}\,,
\end{equation}
where $\x$ is the fermion spectral parameter and $\y$ is the boson
one. This factor appears in the equation on the fermionic node. In
the bosonic-node equation $\x$ and $\y$ are interchanged.
\item{\bf Inversion boson-fermion links:} The inversion links that
connect momentum-carrying bosonic nodes with the "wrong" fermionic
nodes are associated with a factor
\begin{equation}\label{}
 \frac{1-\frac{1}{\x\y^+}}{1-\frac{1}{\x\y^-}}\approx
 1+\frac{i}{h}\,\,\frac{1}{\y^2-1}\,\,\frac{1}{x-\frac{1}{\y}}\,.
\end{equation}
In the equation on the bosonic node $\x$ and $\y$ are again
interchanged.
\end{itemize}
This way one can easily reconstruct the Bethe equations for
$AdS_5\times S^5$ \cite{Beisert:2005fw} from the Dynkin
diagram~\ref{dn224class}; the Bethe equations for $AdS_4\times CP^3$
\cite{Gromov:2008qe} from the diagram~\ref{figdynosp}, and the
conjectured Bethe equations for $AdS_3\times S^3$ or $AdS_3\times
S^3\times S^3$ \cite{Babichenko:2009dk} from the
diagrams~\ref{dps}~and~\ref{dynkinclass}.

\subsection*{Acknowledgments}
This work was
supported in part by the Swedish Research
Council under the contract 621-2007-4177, in part by the ANF-a grant
09-02-91005, and in part by the grant for support of scientific
schools NSH-3036.2008.2.

\bibliographystyle{nb}
\bibliography{refs}

\end{document}